\documentclass[]{aastex631}

\begin{document}

\title{JWST Reveals Varied Origins Between Jupiter's Irregular Satellites}

\author[0000-0003-1383-1578]{Benjamin N. L. Sharkey}
\affiliation{Department of Astronomy, University of Maryland \\
4296 Stadium Dr. \\
PSC (Bldg 415) Rm 1113 \\
College Park, MD 20742-2421, USA \\
}

\author[0000-0002-9939-9976]{Andrew S. Rivkin}
\affiliation{Johns Hopkins University Applied Physics Laboratory, Laurel, MD 20723, USA}

\author[0000-0002-6886-6009]{Richard J. Cartwright}
\affiliation{Johns Hopkins University Applied Physics Laboratory, Laurel, MD 20723, USA}

\author[0000-0002-6117-0164]{Bryan J. Holler}
\affiliation{Space Telescope Science Institute, Baltimore, MD, USA}
\author[0000-0001-9265-9475]{Joshua P. Emery}
\affiliation{Department of Astronomy and Planetary Science, Northern Arizona University, Flagstaff, AZ, USA}

\author[0000-0003-3091-5757]{Cristina Thomas}
\affiliation{Department of Astronomy and Planetary Science, Northern Arizona University, Flagstaff, AZ, USA}
 
\begin{abstract}
We report JWST NIRSpec (0.7-5.1 $\micron$) observations of eight Jovian irregular satellites across five orbital groups. We detect variation in the phyllosilicate content of the the three largest members of the Himalia collisional family (Himalia, $D\sim140km$, Elara, $D\sim80km$, and Lysithea, $D\sim40km$). Himalia contains complexed CO$_2$ and overlapping absorption features from $\sim2.7-3.6\ \mu m$ that match laboratory samples of ammoniated phyllosilicates. Lysithea displays a simpler, single-minimum $3 \micron$ feature caused by an unidentified absorber. Elara presents a 3$\micron$ band that matches by a simple average of Himalia and Lysithea. We argue that the Himalia parent body was heterogeneous and formed with materials similar to Ceres-like ammonium-bearing asteroids, consistent with suggestions from previous visible-wavelength observations. The satellites Carme, Sinope, and Themisto have colors and absorption features similar to ``red'' Jovian Trojans. The satellites Ananke, Pasiphae, and Lysithea each have absorption bands centered from $2.93-2.97\micron$, intermediate between the longer-wavelength bands observed on Trojan asteroids and the shorter-wavelength bands commonly seen on phyllosilicate-rich C2 chondrites. These intermediate-wavelength absorptions are present in both Trojan-like families and the hydrated Himalia family, confounding a link to a single compositional context. The observed differences between Jupiter's irregular satellites and Trojans requires one of two possibilities: 1) some Trojan parent bodies contained hydrated materials in their cores; 2)  some Jovian irregular satellites were not captured from the Trojan parent reservoir. Future searches for NH-bearing materials across the solar system and 3$\micron$ studies of small body collisional families may provide means to discriminate between these two hypotheses.
\end{abstract}

\section{Jupiter's Irregular Satellites: Possible Remnants of Outer Solar System Mixing} \label{sec:intro}

\begin{figure}[h!]
\begin{centering}
\includegraphics[width=0.8\textwidth]{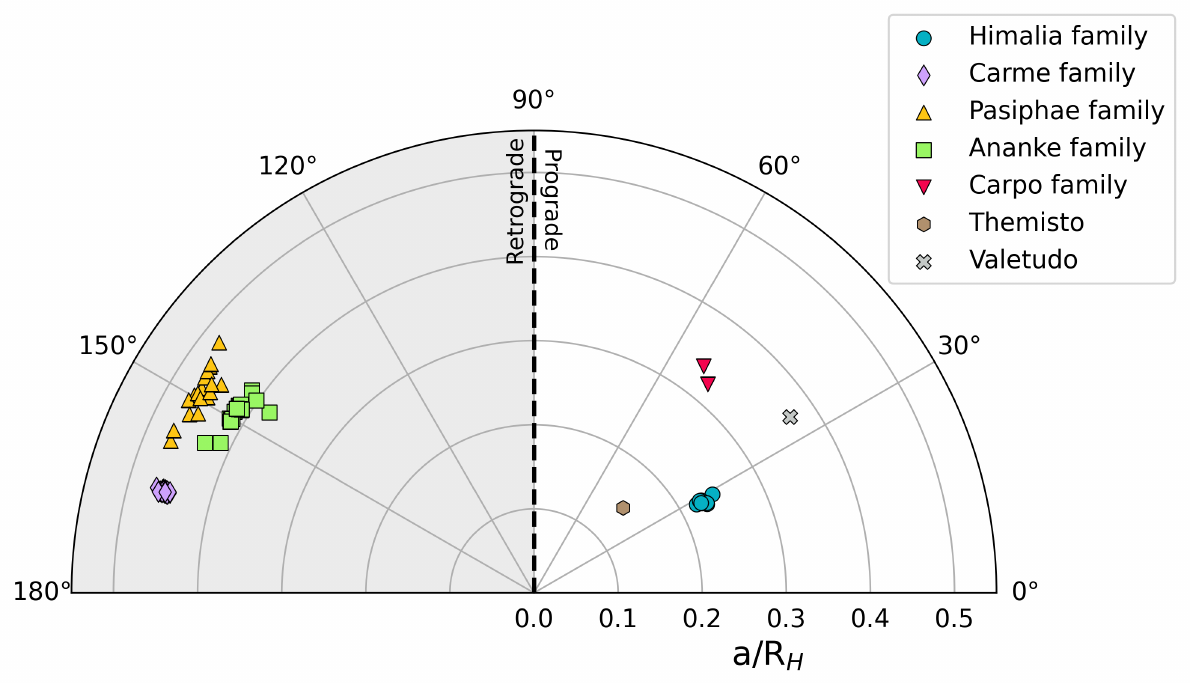}
\caption{Irregular satellite orbital inclinations and semimajor axes (normalized to Jupiter's Hill radius). Family groupings are given as indicated. The prograde population is dominated in mass and number by the Himalia family. The retrograde population is categorized into three orbital families, with the more widely dispersed Pasiphae and Ananke families and the more tightly grouped Carme family. This work studies the Himalia, Carme, Pasiphae, and Ananke families, as well as the isolated satellite Themisto. Orbital parameters from JPL Horizons.}
\end{centering}
\label{fig:orbits}
\end{figure}

The origins of Jupiter's irregular satellites are uncertain.  The highly inclined and eccentric orbits of these small satellites lie exterior to the regular, Galilean satellites, suggesting that they were captured from asteroid- or comet-like parent reservoirs \citep[e.g.,][]{Sheppard2003,Nesvorny2007,Nesvorny2014}. Jovian irregular satellite orbits are also clustered, with one large prograde group (Himalia), three major retrograde groups (Ananke, Carme, and Pasiphae), and several smaller, isolated satellites \citep{Sheppard2003}. The (a,i) orbital elements of the 95 known Jovian irregular satellites \citep{Sheppard2023}, noting major family clusters, are displayed in Figure \ref{fig:orbits}.

Various processes have been proposed to explain the capture of Jupiter's irregular satellites, as well as similar populations at Saturn, Uranus, and Neptune. \citet{Nesvorny2018}, \citet{Jewitt2007}, and \citet{Nicholson2008} provide thorough reviews of these mechanisms. Some relate the capture process to the formation of the planet itself, including pull-down capture \citep[][where the planet's mass growth captures nearby resonant objects]{Heppenheimer1977} and gas-drag \citep[][where nebular gas affects nearby planetesimals that drift into bound states]{Pollack1979,Astakhov2003,Cuk2004,Kortenkamp2005}. Alternatively, three-body interactions could emplace satellites after planet formation \citep{Colombo1971}, including recent suggestions that close stellar flybys could have implanted the irregular satellites \citep{Pfalzner2024}.

Notably, models of planet migration reproduce the overall number and orbital distributions of each giant planet's irregular satellites via three-body interactions between a (to be captured) small body and two closely interacting planets \citep{Nesvorny2007,Nesvorny2014}. Irregular satellite capture may therefore relate to large-scale migratory processes that have sculpted the overall architecture of the solar system. Similar capture models can explain the present-day Jupiter Trojans \citep{Nesvorny2013}, and planet migration provides key insight to understand the main asteroid belt \citep[e.g.,][]{Minton2009,Morbidelli2010,Walsh2011}, Pluto's orbit \citep{Malhotra1993}, and the complex populations in the trans-Neptunian region \citep[see recent review by][]{Gladman2021}.

Through the lens of migratory capture, the most likely source reservoir for irregular satellites and Trojans would be the primordial Kuiper belt. This is a hypothesized region stretching from $\sim24$ and $\sim50$ au \citep{Nesvorny2019}, immediately outside the locations where the giant planets accreted \citep[$\sim5$ and $\sim17$ au, e.g.,][]{Tsiganis2005,Nesvorny2012,Nesvorny2019}. Within $\sim 10-100$ Myr after the dissipation of the solar nebula, Neptune's migration through the primordial Kuiper belt would have destabilized this planetesimal population and triggered dynamic instability throughout the outer solar system \citep{Tsiganis2005}. Planetesimals ejected inwards would enter planet-crossing orbits, and those ejected outwards would create the scattered disk \citep{Levison2008}. Meanwhile, the present-day dynamically ``cold" Kuiper belt population would represent the portion of the primordial Kuiper belt that formed far enough away from the giant planets to remain stable \citep[although this population may have also been largely depleted in mass, see e.g.][]{Gomes2018}

While the irregular satellites could have been captured from a single source, they could also represent a composite of various populations that interacted with Jupiter over the planet's history. The three-body capture scenarios defined by \citet{Nesvorny2007} occur in the vicinity of primordial Kuiper belt remnants, not the original generation of middle solar system planetesimals. However, some closely-bound irregular satellites may be able to survive from previous capture events. For example, \citet{Nesvorny2014} find that $\sim 1-30 \%$ of Jovian satellites with closer-in, Himalia-like orbits could survive planetary encounters that occurred after their capture. The likelihood of this successive capture scenario is currently speculative, but could be constrained by determining whether or not all irregular satellite parent bodies formed within similar primordial environments.

Observational evidence does not yet uniquely discriminate irregular satellites' formation region(s). Jupiter's position, proximate to both the inner solar system's reservoirs of asteroids as well as the outer solar system's comets, trans-Neptunian objects (TNOs), and Centaurs, provides a rich array of comparison populations. Determining the irregular satellites' compositions in the context of these other populations tests how migratory capture could have occurred. But the post-capture evolution of the irregular satellites, including surface alteration and collisional breakup, must be accounted for in order to forge successful cross-population links. To determine the lineage(s) of Jupiter's captured satellites, we therefore require knowledge of their bulk compositions as well as the variability within various collisional families. We provide these bulk compositional assessments with high-SNR JWST spectra of the largest objects in each major satellite family, and we provide the first detailed exploration of $3 \micron$ variability within a collisional family as a function of size. These constraints address irregular satellite formation conditions and the evolutionary processes that have shaped them.

\subsection{Previous Observational and Laboratory Context}

Numerous spacecraft missions have explored the Jovian system, but only the largest irregular satellite, Himalia, has been characterized by spacecraft imaging (resolved across $\sim$5 pixels by \citealt{Porco2003}, unresolved by \citealt{Cheng2010}) and spectroscopy \citep[][SNR$\sim 10$]{Brown2003,Chamberlain2004}. \citet{Chamberlain2004} reported the ``suggestion of an absorption'' near 3 $\micron$ on Himalia, but additional spectral structure could not be retrieved from those data. Telescopic observations have suggested kinship between irregular satellites and various groups of dark primitive asteroids. \citet{Jarvis2000} reported a 0.7 $\micron$ feature on Himalia, suggesting that Himalia contains phyllosilicates and is taxonomically similar to the main belt asteroid (52) Europa. \citet{Brown2014} detected and characterized Himalia's 3$\micron$ band, reporting this feature to also match (52) Europa \citep{Takir2012} but not match any known phyllosilicates. This feature is only partially observable from ground-based telescopes because of atmospheric opacity in the $\sim2.5-2.85-\micron$ region, which limits compositional tests by obscuring diagnostic information like the locations of band centers. Himalia and its associated family were also reported to display a 0.7 $\micron$ feature of variable strength correlated with size by \citet{Vilas2024}, who also report a similar feature in some Ananke and Pasiphae spectra.  Phyllosilicates are also components of the best spectral fits to Himalia and Elara by \citet{Bhatt2017}.


Amongst other satellite families, numerous neutral-to-very-red spectra have been observed, similar in color to Jupiter Trojan asteroids and without apparent absorption features at wavelengths shorter than $2.5 \micron$ \citep{Sharkey2023,Bhatt2017}. Coarser color photometry surveys \citep{Graykowski2018,Grav2003,Grav2004a} have characterized a wider satellite sample and have found Jupiter's irregular satellites to have colors similar to those of Saturn, Uranus, and Neptune. But the relative scarcity and faintness of Jupiter's irregular satellites have limited further attempts to characterize them. These observational constraints generally limit even photometric studies to the few Jovian satellites that are larger than D $\sim$ 10 km, and they have precluded the high-SNR spectroscopy that is needed to interpret near-infrared mineral absorption features.

Existing high-resolution, high-SNR spectra of Galilean satellites may provide an alternate means to investigate irregular satellite compositions. Collisions within irregular satellite swarms generate dust that likely collides with the surfaces of the Galilean moons, in particular the leading hemisphere of Callisto \citep[e.g.,][]{Bottke2010,Bottke2013,Bottke2024,chen2024}. Accumulated dust from these primordial bodies then mixes with native components in the Galilean moons' regoliths that is subsequently irradiated by charged particles trapped in Jupiter's massive magnetosphere \citep{johnson2004}. This process would form new components, such as CO$_2$ \citep[e.g.,][]{gomis2005}, sulfates \citep[e.g.,][]{carlson2002}, and perhaps complex CN-bearing organics \citep[e.g.,][]{cartwright2024} depending on the nature of the exogenic material. 

However, the lack of high quality spectral datasets for Jupiter's irregular satellites limits the scope of hypothesis testing both for irregular satellite compositions and their connections to the Galilean moons. In particular, a broad band centered near 4.57 $\micron$ on Callisto's leading hemisphere has been speculated to have an irregular satellite origin \citep{cartwright2020,cartwright2024}, but this speculation cannot be tested without searching for the same feature on irregular satellites. Without identifying specific classes of silicate minerals, ices, or organic species, the origins of the irregular satellites, and their effects on the Jovian system, remain mysterious.

Laboratory studies of meteorites further reinforce the need to characterize absorption features to properly identify minerals, ices, or organics on irregular satellites. Petrologic, isotopic, and mineralogical study of the meteorite Tarda suggests that it is a sample from a parent body that is either the same or closely related to that of the Tagish Lake meteorite. This link is particularly interesting given that Tarda's spectrum is ``less-red" while Tagish Lake is ``red" \citep{Schrader2024}. More generally, a variety of carbonaceous chondrites with wildly different alteration histories and compositions (petrographic grades 1-3) can have remarkably similar colors from $\sim$ 0.7-2.5 $\micron$ \citep[e.g.,][]{Cloutis2011a,Cloutis2011b}. Strikingly, the color variation within individual sample preparations of some CM and CI chondrite samples can exceed the differences between C/P/D-type asteroids (see packing conditions of various spectra of CI Alais in \citealt{Cloutis2011a}; see grain-size variations in both CM2 Aguas Zarcas and Tarda from \citealt{Cantillo2021} and CM2 Murchison by \citealt{Cloutis2018}). Given the magnitude of these noncompositional spectral effects, \citet{Sharkey2023} noted that differences in the activity histories between red/less-red objects (speculatively caused by different volatile abundances) could be a confounding driver of their color differences, even if their present-day surfaces contain similar materials.

With the advent of JWST, our understanding of small solar system bodies is rapidly expanding. JWST's NIRSpec instrument's high sensitivity over key infrared passbands identifies assemblages of rocky minerals, volatile ices, and organic compounds with spectroscopic complexity previously accessible only via laboratory study.

Recent observations of Jovian Trojans by \citet{Wong2024} highlight the importance of studying carbonaceous asteroids from 0.7-5.2 $\micron$. When observed from Earth-based telescopes, Trojans are featureless $<$ 2.5 $\micron$, and the presence of the 3 $\micron$ bands were partially detectable \citep{Brown2016} but not fully characterized due to telluric contamination. When observed via JWST, distinct spectral features from 2.5-5.0 $\micron$ in both color groups were characterized in full, providing clear identifications of organic compounds and OH-bearing materials on these asteroids. 

The \citet{Wong2024} characterization of 3$\micron$ bands on Trojans provides new insight into the evolution and origins of red/less-red surfaces. If Jovian Trojans formed originally in the primordial Kuiper belt, they would have lost any surface ices that become volatile between 5-40 au during their inward migration. But more complex processing could have also occurred to produce stable compounds from volatile mixtures, such as hypothesized space weathering effects \citep[e.g.,][]{Wong2016}. The irradiation of organic and ice-rich materials have been demonstrated to create complex assemblages in laboratory simulations of asteroidal surfaces \citep{Urso2020}. Such scenarios have been hypothesized to explain the rounded 3 $\micron$ absorption features on some main-belt asteroids, suggesting that they may have formed in outer solar system reservoirs under conditions cold enough to incorporate methanol ice and ammonium \citep{Rivkin2022}. This hypothesis may also apply to main-belt comets like 238P/Read, which has a similarly rounded $3 \micron$ band \citep{Kelley2023}, as well as Jupiter Trojan asteroids with similarly shaped $3 \micron$ absorptions \citep{Wong2024}.

Coupled with their shared heliocentric distance, the strong spectral slope similarities between the Jovian Trojans and irregular satellites already suggest some level of compositional similarity. However, the effects of more frequent collisions within the irregular satellites \citep{Bottke2010}, and the presence of objects like Himalia with absorption features that are comparable to outer-main belt asteroids \citep{Jarvis2000,Brown2014,Takir2012,Vilas2024} complicates a direct equivalence between the two populations. This work further constrains the mineralogies of Jupiter's irregular satellite groups, shedding new light on the organic-rich material reservoirs that have interacted with Jupiter and the Galilean moons.

\section{Observations and Data Reduction} \label{sec:obsdata}

\begin{table}[h!]
\begin{center}
\begin{tabular}{|c|c|c|c|c|c|c|c|c|}
\hline \hline
\textbf{Satellite} & Family &Diameter (km) & Observation Date & Phase Angle & Sun-Target & Target-Observer \\
 & & & & (Degrees) & Distance (au) & Distance (au) \\
\hline
 JXVIII Themisto & Themisto & 8\tablenotemark{a} & 2024-01-31 & 11.59 & 4.94 & 4.90 \\
\hline
JVI Himalia & Himalia & $139.6\pm 1.7$\tablenotemark{b} & 2024-01-31 & 11.54 & 4.96 & 4.92 \\
\hline
JX Lysithea & Himalia & $42.2\pm0.7$\tablenotemark{b} & 2024-01-25 & 11.55 & 4.96 & 4.82 \\
\hline
JVII Elara & Himalia & $79.9\pm1.7$\tablenotemark{b} & 2024-01-18 & 11.40 & 4.97 & 4.72 \\
\hline
JXII Ananke & Ananke & $29.1\pm0.6$\tablenotemark{b} & 2024-01-31 & 11.19 & 5.11 & 5.08 \\
\hline
JXI Carme & Carme & $46.7\pm0.9$\tablenotemark{b} & 2024-01-25 & 11.92 & 4.81 & 4.69 \\
\hline
JVIII Pasiphae & Pasiphae & $57.8\pm0.8$\tablenotemark{b} & 2024-01-25 & 11.53 & 4.97 & 4.89 \\
\hline
JIX Sinope & Pasiphae & $35.0\pm0.6$\tablenotemark{b} & 2024-01-21 & 11.58 & 4.94 & 4.76 \\
\hline
\end{tabular}
\tablenotetext{a}{Via \citet{Sheppard2003}, assumes geometric albedo of 0.04.} 
\tablenotetext{b}{Via \citep{Grav2015} from NEOWISE thermal-IR observations.} 
\end{center}
\caption{Dates of reported NIRSpec observations, including relevant observing geometry for computing thermal model fits to correct spectral irradiance to reflectance. Geometry reported via JPL Horizons. Ordered by satellite semimajor axis.}
\label{tab:circumstances}
\end{table}

Observations were performed using the NIRSpec integral field unit (IFU) in prism mode ($R\sim100$) for all targets except Himalia, which was observed in the higher resolution G235M and G395M grating modes ($R\sim1000$)\footnote{Grating-specific dispersion and spectral resolution files are available at \href{https://jwst-docs.stsci.edu/jwst-near-infrared-spectrograph/nirspec-instrumentation/nirspec-dispersers-and-filters\#gsc.tab=0}{NIRSpec Dispersers and Filters} in the JWST documentation.}. Observational circumstances, including timing and geometry, are given in Table \ref{tab:circumstances}.

Calibration and spectral extraction of the Jovian irregular satellite data proceeded similarly to the processes described in \citet{SouzaFeliciano2024,DePra2024,Licandro2024,Pinilla-Alonso2024}. This process is similar to that described in \citet{Wong2024}, with some differences. To start, \textit{uncal} files for all targets were downloaded from the Barbara A. Mikulski Archive for Space Telescopes (MAST) using the \textit{jwst\_mast\_query} tool and processed through the JWST data calibration pipeline locally. Version 1.14.0 of the calibration pipeline \citep{Bushouse2024} and reference context \textit{jwst\_1225.pmap} were used. The pipeline was run with default parameters and included the non-default NSClean step to remove the 1/\textit{f} pattern noise \citep{Rauscher2024}. Processing was performed up to Level 2b and the production of dither-specific \textit{s3d} spectral data cubes.

Spectral extraction on each individual data cube made use of the ``template PSF-fitting" routine first described in \citet{Wong2024}. The uniformly spaced wavelength array was constructed and slices $<$0.7 $\mu$m and $>$5.1 $\mu$m were trimmed due to higher noise. Then, the data quality (DQ) extension was used to identify pixels flagged by the pipeline and any pixel with a non-zero value was set to NaN for the remainder of the extraction. Target centroids were first determined by-eye (typically good to a few tenths of a pixel) and used to measure radial distances from the target. Outlier pixels due to artifacts or background objects, as well as those $>$5 pixels from the centroid were set to NaN. Then the background was calculated in each slice of the cube using the median of all pixels $>$5 pixels from the centroid and subtracted from that slice. A template PSF was then computed as the median of the 10 background-subtracted slices proceeding and succeeding the slice under consideration (inclusive) for a total of 21 slices. Slices near the edge of the considered wavelength range included $<$21 total slices, but the resulting template PSFs were still adequate. The template PSF was then trimmed to a 9$\times$9-pixel vignette, with all external pixels set to NaN, in order to reduce the leverage of more distant pixels when fitting to the central slice in the set. The integrated flux of the vignette was then normalized to unity. The \textit{scipy.optimize.minimize} function and a Nelder-Mead (amoeba) algorithm was used to determine the best-fit values for the flux scaling factor of the template PSF and the background level. This fit was iteratively performed 3 times, using the outputs from the previous iteration as the initial guess for the next iteration. Additionally, after each iteration a model was constructed (template PSF multiplied by the flux scaling factor followed by the addition of the background) and subtracted from the data, with any pixels $>$5-$\sigma$ from the mean counted as outliers and set to NaN prior to the next iteration. After the third iteration, a center-of-mass centroid was computed to determine the target centroid to sub-pixel accuracy, the template PSF was multiplied by the flux scaling factor, and the flux was extracted within a 3.5-pixel radius circular aperture.

Following the construction of the 1D spectra as described above, the individual dithers were combined and the reflected solar component was divided out. The G-type standard star SNAP-2 (Gordon et al. 2022), observed in program 1128 (PI: Lutzgendorf), was used for this purpose and these data cubes underwent the same processing as described above for the irregular satellites data cubes. All individual standard star spectra were first resampled onto the same wavelength grid and normalized by the median of all dithers. The purpose of this resampling and normalization was to reduce the spurious spread in absolute flux values due to placement of the target at different locations within the IFU field of view. The median was then computed in each wavelength bin and multiplied by the previously calculated median to retrieve the absolute flux scaling. Three iterations of outlier rejection were performed with a 21-point moving median and a 3-$\sigma$ threshold; any points found to exceed this threshold were replaced with the value of the moving median. The irregular satellite 1D spectra underwent the same resampling, normalization, median, rescaling, and outlier rejection steps and then were divided by the SNAP-2 spectrum. Uncertainties on each medianed spectrum were computed as the median absolute deviation of each wavelength bin and uncertainties were propagated following division by the standard star.

\begin{figure}[h!]
\begin{centering}
\includegraphics[width=0.8\textwidth]{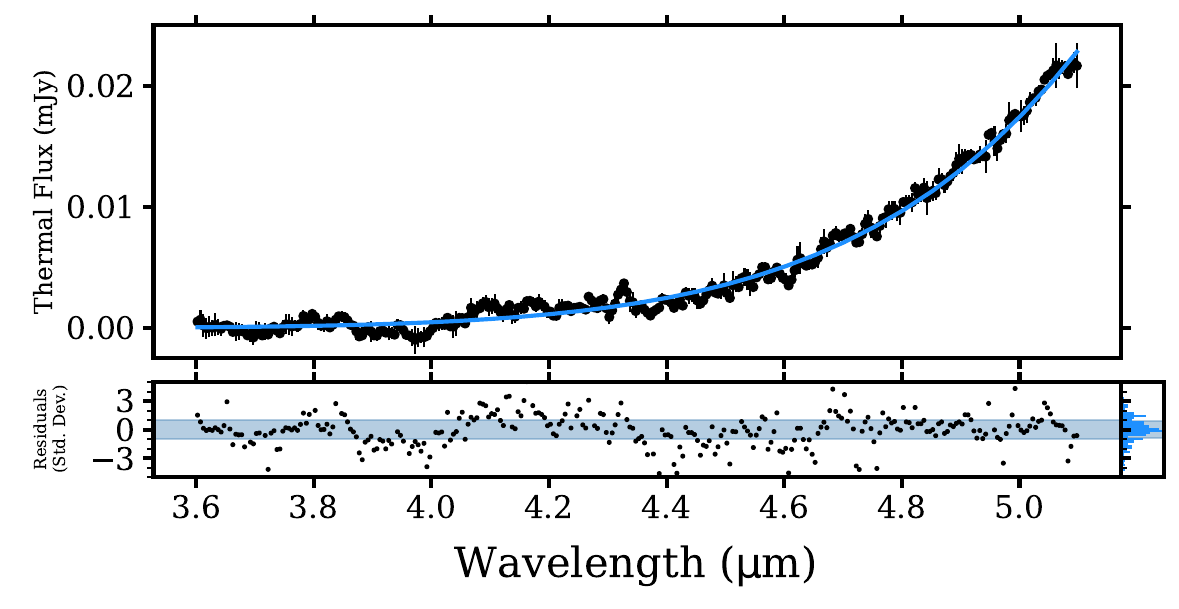}
\caption{Top: Estimated thermal flux (data points) and resulting NEATM best-fit thermal model (blue curve) for irregular satellite Lysithea. Fits were performed by varying target albedo and beaming parameter to reproduce the cut-on wavelength where measurable thermal emission begins (typically near 4.0 microns, related to the observed surface's temperature), as well as the shape and magnitude of the thermal fluxes at longer wavelengths. Finalized reflectance spectra are created by subtraction of the blue curve from the processed, flux-calibrated spectra. Bottom: Thermal flux residuals from best-fit NEATM Model.}
\end{centering}
\label{fig:neatm_example}
\end{figure}

Our processed, flux calibrated spectra include a component of thermal emission that begins near 4.0 microns and increases in strength to dominate observed fluxes $>$ 4.5 microns. The thermal emission component manifests as a broad, blackbody-like curve that distorts our analysis of reflected light. To isolate the reflected component of the observed flux, we estimated the thermal component via an iterative implementation of the near-Earth Asteroid Thermal Model \citep[NEATM,][]{Harris1998,Delbo2002}. This model estimates thermal flux from an asteroidal surface by accounting for observing geometry, asteroid size (diameter), albedo, and a beaming parameter, $\eta$, which broadly accounts for variations according to different surface-scattering scenarios. For the purposes of our spectral corrections, we took the geometric albedo and $\eta$ as free parameters and took values of satellite diameters from \citet{Grav2015}.

The NEATM models were fit using \textit{emcee}, a Markov-Chain Monte Carlo package in python \citep{foremanmackey2013}. We took albedo values of 2\% and $\eta$ values of 1.0 as our initial solutions for all objects. Our goals for these models are not to provide detailed explorations of thermal model parameters (including their possible degeneracies), but instead to estimate the shape of the thermal flux curve. To do so, some assumption about the shape of the underlying reflectance spectra is required. We estimated the reflected component of each satellite spectrum as a linear extension of the observed spectral slope at the cut-on point of the thermal emission (variable, near 4 microns). MCMC fits were then performed to recover NEATM model solutions and best-fit parameters. Specifically, fits were iterated to ensure good agreement to the cut-on wavelengths and the magnitudes of peak thermal emission. The choice of baseline was iterated until a smooth output spectrum was produced, to ensure that spectral slopes were not extrapolated from regions with absorption bands. Figure \ref{fig:neatm_example} shows a representative example of a finalized thermal flux estimate and the NEATM best-fit model for Lysithea. Reflectance spectra were then computed by subtracting the NEATM best-fits from the original flux-calibrated spectra. 

\section{Spectral Groupings and Band Identifications} \label{sec:results}

\begin{figure}[h!]
\includegraphics[width=.43\textwidth]{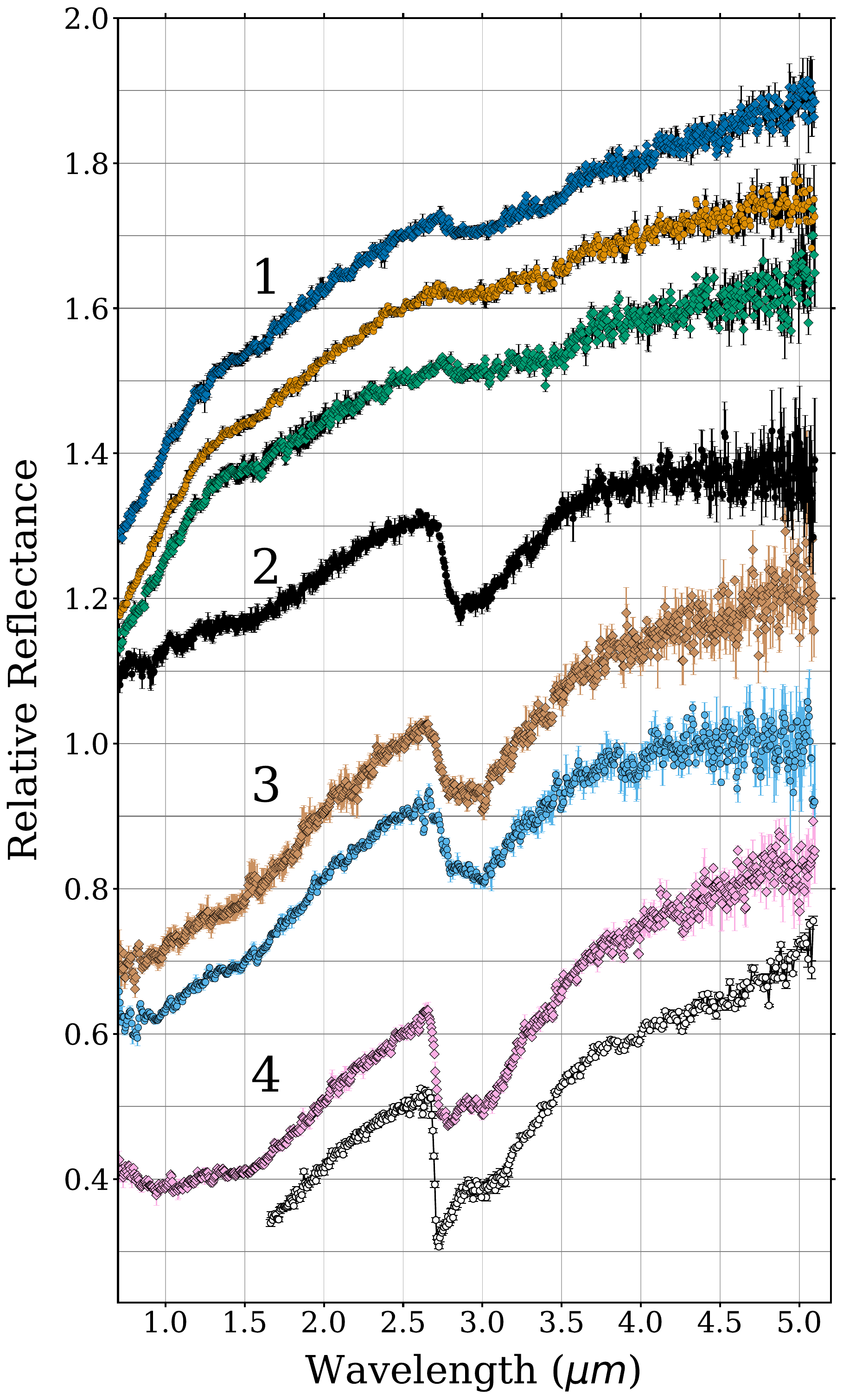}
\includegraphics[width=.565\textwidth]{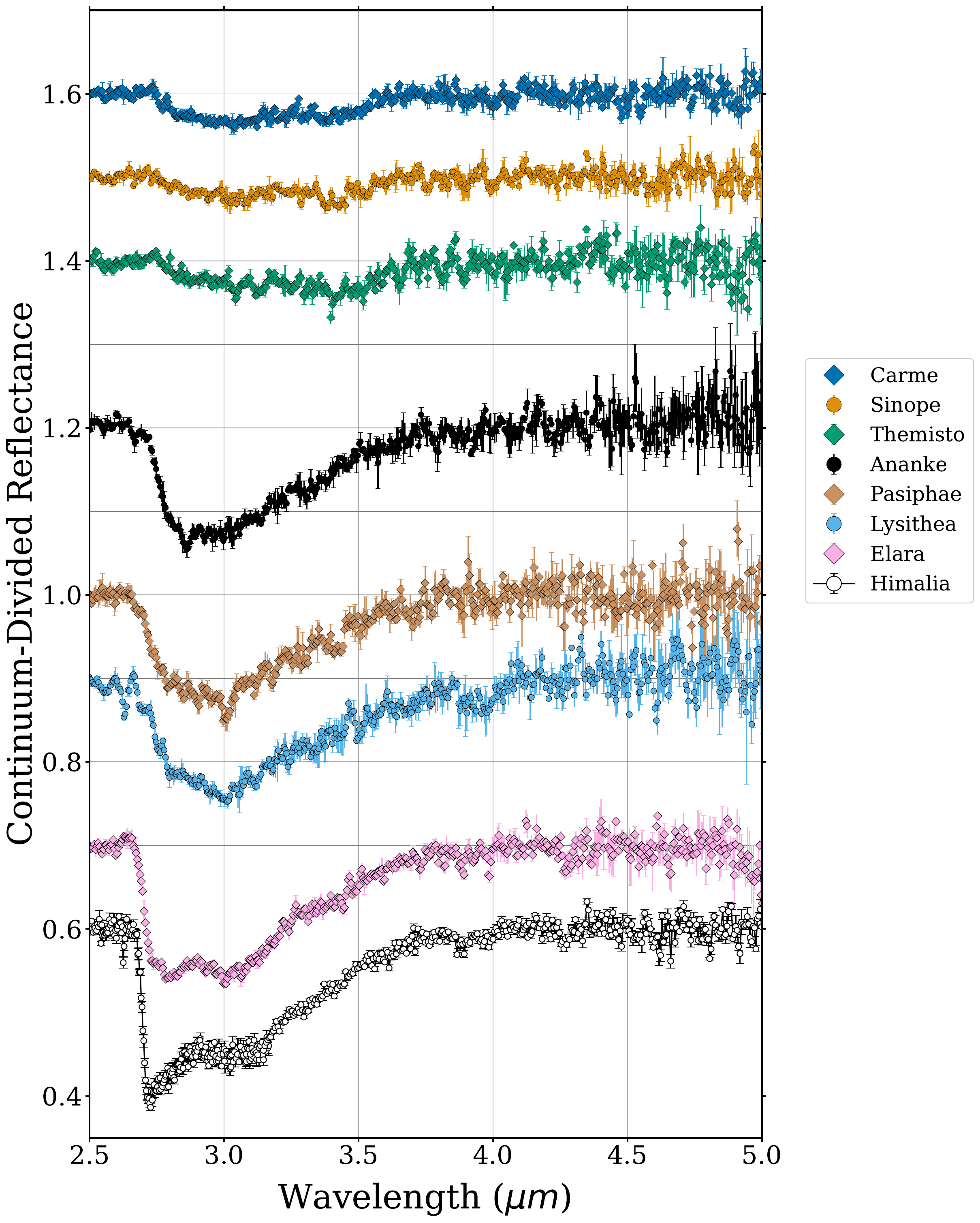}

\caption{Left: Processed relative reflectances for all observed targets in JWST GO 4028, labeled according to groupings identified in the text. }Right: Continuum-divided reflectances using a polynomial of order 2-4 fit on either side of major absorption bands. Objects are normalized to have reflectance equal to 1 at 2.5 microns, and displayed with offsets for ease of comparison. Targets are arranged according to spectral characteristics, not orbital groupings. Major band complexes near 3.0 microns differentiate between the different material categories. The steep red slopes of Carme, Sinope, and Themisto (Group 1) are similar to D-type asteroids or “red” type Jovian Trojans. Ananke (Group 2) shows a broad feature at similar wavelengths to Lysithea and Pasiphae but with a very different, rounded shape. Lysithea and Pasiphae (Group 3) show deep features centered on 3.0 microns similar to less-red Jovian Trojans. Himalia and Elara (Group 4) display a clear 2.7 micron feature, detected previously on aqueously altered inner-solar system materials but never identified in the Jovian system.
\label{fig:alldata}
\end{figure}

\subsection{Continua Definitions}
The processed, thermally-corrected reflectance spectrum of each target is shown in Figure \ref{fig:alldata} (Left). Proper analysis of absorption features requires clear definitions of the spectral continuum. To aid reader comparisons, we define a ``global'' continuum, used in constructing Figure \ref{fig:alldata} (Right). This continuum was fit per-object with a low order (2-4) polynomial on either side of the identified 3 micron complex, typically near 2.5 and 4.5 $\micron$. Dividing the spectrum by this polynomial transforms the spectra from relative reflectance to continuum-divided reflectances. In continuum-divided space, spectral features are identified via strong divergence from unity.

This global continuum is well-suited to identify major features, but it is imprecise for characterizing weak and/or narrow absorption bands. We report band centers and depths in Table \ref{tab:BandProperties} by defining local continua as follows. For the broad 3 $\micron$ band, a linear continuum was fit from $\sim$2.7 to 3.8 microns, with the centers then identified by a polynomial (3rd degree) fit to the bottom half of the band. The listed depth is from this linear continuum to the center position. We additionally report the band depth at 2.9 $\micron$ to facilitate comparisons with ground-based observations.

The edges of all other bands were identified by eye, and continuum points were defined as the average within a 0.01 um box on either side of the feature.  A linear continuum was fit to these points and divided out, then a Gaussian was fit to identify the center and depth from the local continuum. Due to its asymmetry, we quantify Himalia's 2.7 $\micron$ feature uniquely. Its center is defined by fitting a Gaussian near its sharp minimum, and its depth is defined relative to the linear continuum of the overall 3 $\micron$ band complex.

To determine the robustness of individual features, band parameters are reported according to a Monte Carlo resampling procedure. After continuum division, individual data points are resampled according to their Gaussian uncertainties before fitting feature depths and centers. Quoted parameters are the mean of 10,000 resampling iterations, and the standard deviations of the resampled parameters taken as the uncertainty.

\subsection{Spectral Groups}

We find that the irregular satellites can be roughly described by four groups according to their spectral slopes and 3$\micron$ absorption features. These groupings include some that appear only within individual families, and some that appear in multiple families. Band parameters for all features observed on individual satellites are given in Table \ref{tab:BandProperties}. The first group is represented by Sinope, Carme, and Themisto, which are all red-Trojan-like objects and are all in different orbital families. Ananke comprises the second group, displaying a deeper, rounded 3.0 micron band and relatively neutral spectral slopes. The third group includes Pasiphae, a retrograde satellite, and Lysithea, a prograde satellite, which have 3.0 $\micron$ features centered similarly to Ananke, but with redder spectral slopes and a sharper shape from 2.7-3.3 $\micron$. Finally, Himalia and Elara, both in the prograde Himalia family, show a unique band complex that includes a previously unobserved sharp 2.7 micron band that transitions into the broad, round feature from 2.8-3.5 microns that was observed by \citet{Brown2014}.

Sinope and Carme have similarly steeply increasing spectra from 0.7-2.5 microns, and shallow absorption features near 3.0 and 3.4 $\micron$, akin to the ``red" class of Jupiter Trojans \citep{emery2011,Wong2024}. Notably, Themisto, a small ($D<10\ km$) individual prograde satellite interior to the Himalia family, groups closely with these more weakly bound retrograde satellites. The spectral slopes of Sinope and Carme, when overlapping, are consistent with previous measurements at VNIR wavelengths by \citet{Sharkey2023}. Carme has a slightly deeper band at 3.0 microns than Sinope and Themisto (depths of $\sim 3\%$ vs. $\sim2\%$), but they all have similar 3.4 $\micron$ depths ($\sim2\%$). 

\begin{table}[h!]
\begin{center}
\caption{Derived band measurements, including compositional attributions. }
\resizebox{\textwidth}{!}{%
\begin{tabular}{|c|c|c|c|c|c|c|c|c|c|}
\hline \hline
& & Carme & Sinope & Themisto & Ananke & Pasiphae & Lysithea & Elara & Himalia \\
\hline \hline
                 & Center ($\mu m$) & - & - & - & $2.67 \pm 0.03 $& - & $2.63 \pm 0.01$ & $2.60 \pm 0.07$ & $2.63 \pm.01$ \\
\cline{2-10}
$2.63 \mu m$ Band & Depth (\%) & - & - & - & $1.4 \pm 1.4$ & - & $4.63\pm0.4$ & $1.2\pm 1.5$& $1.4 \pm2.9$ \\
\cline{2-10}
                 &  Attribution & - & - & - & ? & - & Carbonates? & ? & ? \\
\hline
\hline
                 & Center ($\mu m$) & - & - & - & $2.861\pm0.003$ & $3.010\pm 0.002$& $3.00 \pm 0.01$ & $2.771 \pm 0.002$ & $2.719 \pm 0.002$ \\
\cline{2-10}
$2.7-3.0 \mu m$& Depth (\%) & - & - & - & $2.7\pm0.5$ &$4.2\pm0.6$ & $2.0\pm0.9$& $12.4\pm0.7$ & $18.67\pm0.23$\tablenotemark{a}\\
\cline{2-10}
 Sharp Band      &  Attribution & - & - & - & ? & ? & ? & Mg-rich & Mg-rich\\
                 &  & & & & & & & Phyllosilicate & Phyllosilicate\\
\hline
\hline
                 & Center ($\mu m$) & $3.01\pm0.02$ &$3.05\pm0.02$ & $3.08\pm0.03$ & $2.93\pm0.005$ & $2.94\pm0.01$& $2.97\pm0.01$& $3.006\pm0.003$ & $3.024 \pm 0.005$ \\
\cline{2-10}
$3.0 \mu m$ Band & Depth (\%) & $2.9\pm0.2$  & $1.9\pm0.2$ & $1.7\pm 0.2$ & $12.0\pm0.2$ & $10.1\pm0.3$ & $10.8\pm0.3$ & $13.4\pm0.2$ & $13.9\pm0.1$ \\
\cline{2-10}
                 &  2.9$\mu m$ Depth & $2.4\pm0.2$ & $1.3\pm0.1$ & $1.6\pm0.2$ & $12.0\pm0.2$ & $10.2\pm0.3$ & $9.4\pm0.2$ & $11.8\pm0.2$ & $13.2\pm0.2$ \\
\cline{2-10}
\hline
\hline
                 & Center ($\mu m$) & $3.42\pm0.01$& $3.412\pm0.004$\tablenotemark{b} & $3.43\pm0.02$ & $3.34\pm0.02$ & $3.42\pm0.01$ & - & $3.41\pm0.02$ & - \\
\cline{2-10}
$3.4 \mu m$ Band & Depth (\%) & $1.8\pm0.3$ & $2.1\pm0.2$ & $2.3\pm0.4$ & $1.8\pm0.4$ & $2.9\pm0.3$ & - & $0.9\pm0.5$ & - \\
\cline{2-10}
                 &  Attribution & Organics & Organics & Organics & Organics & Organics & - & Organics? & - \\
                 &              & (Aliphatic) & (Aliphatic) & (Aliphatic) & (Aromatic) & (Aliphatic) & - & (Aliphatic) & - \\
\hline
\hline
                 & Center ($\mu m$) & - & - & - & - & - & $3.96\pm0.01$ & $3.91\pm0.02$ & $3.94\pm0.01$ \\
\cline{2-10}
$3.9 \mu m$ Band & Depth (\%) & - & - & - & - & - & $2.5\pm0.3$ & $1.7\pm0.7$ & $1.7\pm0.4$ \\
\cline{2-10}
                 & Attribution & - & - & - & - & - & ? & Carbonates? & ? \\
\hline
\hline
                 & Center ($\mu m$) & - & - & - & - & - & - & $4.273\pm0.004$ & $4.268\pm0.002$ \\
\cline{2-10}
$4.27 \mu m$ Band & Depth (\%) & - & - & - & - & - & - & $2.8\pm0.7$ & $2.2\pm0.2$ \\
\cline{2-10}
                 &  Attribution & - & - & - & - & - & - & Complexed & Complexed \\
                 &              & - & - & - & - & - & - & $CO_2$ & $CO_2$ \\
\hline
\hline
                 & Center ($\mu m$) & $4.92\pm0.01$ &$4.90\pm0.03$ & $4.91\pm0.01$ & - & - & - & - & - \\
\cline{2-10}
$4.9 \mu m$ Band & Depth (\%) & $3.3\pm0.7$ & $2.4\pm1.4$ & $5.2\pm1.2$ & - & - & - & - & - \\
\cline{2-10}
                 &  Attribution & ? & ? & ? & - & - & - & - & - \\
                 \hline
\end{tabular}}
\label{tab:BandProperties}
\end{center}
\tablenotetext{a}{Himalia band depth is from the top continuum across the 3$\micron$ band, not local. All others are local.}
\tablenotetext{b}{Sinope organics band may have two components – $3.403\pm0.005 \micron (2.2\pm 0.3\%)$ and $3.527\pm0.005 \micron (1.02\pm1.23\%$). Values in the table assume a single band.}
\end{table}

We also identify possible absorption features at 4.9 $\micron$ on Carme, Sinope, and Themisto. These features are within lower-SNR spectral regions, which are affected by thermal tail removal that  impacts the precision of continuum definitions. If real, this feature is similar to one observed by JWST on the surface of Callisto \citep{cartwright2024}. Its origin is speculative, but could relate to radiolytic production of $CO_3$ from $CO_2$, or the C-O symmetric stretching mode ($\nu_1$) of carbonyl sulfide (OCS). OCS is a common sulfur bearing species observed in interestellar ices in dense molecular clouds \citep{Palumbo1997,Ferrante2008,McClure2023} and cometary comae \citep{Saki2020,Biver2019}.

Ananke has the most neutral (least-red) spectral slope within our sample and a deep, rounded 3.0 $\micron$ absorption band. Its 3 $\micron$ band depth ($\sim 12\%$) and center ($\sim 2.95 \micron$) is similar to the redder Pasiphae and Lysithea, though some distinctions in band shape are present. A narrow feature at 2.86 $\micron$ is overprinted within Ananke's rounded 3 $\micron$ complex. Beyond 3.6 $\micron$, Ananke's spectrum is neutral and featureless. Our observations of Ananke differ from the ground-based observations of \citet{Sharkey2023} from 0.7-1.6 microns, which showed a slope that is approximately 10\% redder. A similar discrepancy was noted by \citet{Wong2024} when comparing their JWST observations of the Trojan Polymele to \citet{Sharkey2019}'s ground-based observations with the same methodology and setup. Given that both Polymele and Ananke have similar brightnesses V $\sim$ 19, it is possible that those short wavelength slope discrepencies are indicative of systematic errors in the ground-based observations of these very faint objects. Previous study from \citet{marsset2020} characterized the spectral slope uncertainties for similar ground-based observations to be $\sim 4\%/\micron$.

Pasiphae (Pasiphae family) groups closely with Lysithea (Himalia family) in terms of spectral slope and absorption features. They trace closely together shortward of $3.6 \mu m$. Pasiphae displays a steeper red slope from 3.5 to 5.2 $\micron$, while Lysithea rolls over to neutral (flat) reflectance, similar to Ananke. Pasiphae and Lysithea display deep $3 \micron$ absorption features ($\sim 10 \%$) with similar positions to Ananke's. 

\begin{figure}[t!]
\begin{centering}
\includegraphics[width=0.75\textwidth]{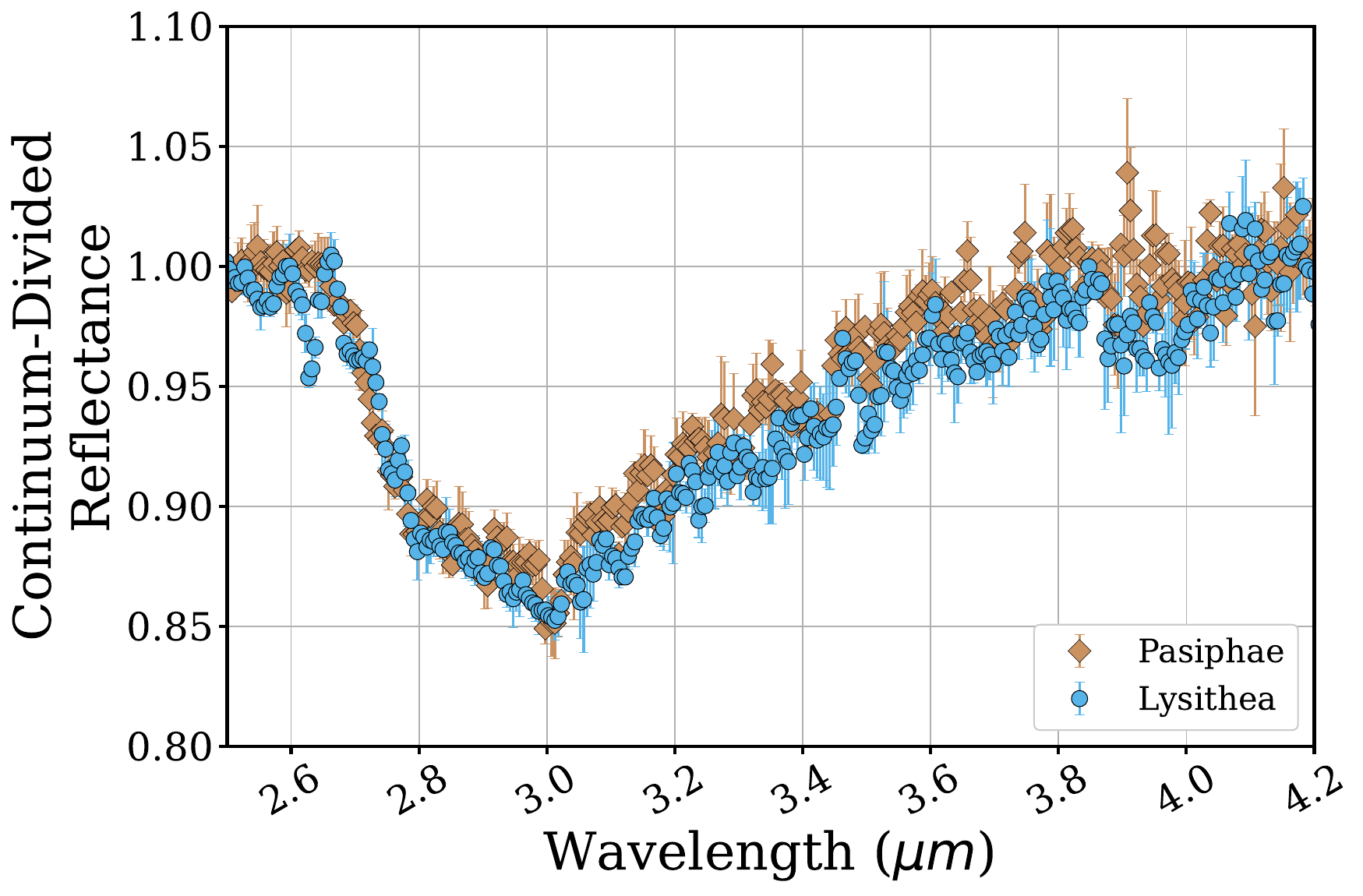}
\caption{Comparison of the 3 $\micron$ band complex on Lysithea (prograde, Himalia group) and Pasiphae (retrograde, Pasiphae group). The centers and depths of these features are highly similar. Lysithea includes an additional narrow absorption near 2.63 $\micron$, and Pasiphae includes a narrow absorption at 3.01 $\micron$ which may also be present on Lysithea.}
\label{fig:ly_pas}
\end{centering}
\end{figure}

A comparison of Pasiphae's and Lysithea's continuum-divided  $3\micron$ bands are shown in figure \ref{fig:ly_pas}, highlighting their similar absorption shapes. Both Pasiphae and Lysithea have additional faint features of 3\% overprinted near 3.01 $\micron$. The cause(s) for the major 3 $\micron$ absorptions on Pasiphae and Lysithea are not clear. Lysithea has an additional feature near 3.9 $\micron$, which is similar to features caused by carbonates. However, carbonates also display absorptions near 3.4 $\micron$ which we do not detect, confusing a clear identification. Lysithea also shows a narrow feature at 2.63 $\micron$ whose origin is also unclear, although it is near trace features seen in some carbonates \citep{Bishop2021}.

Himalia and Elara, both in the prograde Himalia family, are largely similar except for their 2.7$\micron$ bands. They have similar spectral slopes from 1.6-5.2 $\micron$, and similarly located major absorption features at 2.7 and 3.05 $\micron$. However, the relative strength of those major absorption features differs strongly, with Himalia showing a much deeper 2.7$\micron$ minimum ($18.7\pm0.2\%$) than Elara ($12.4\pm0.7\%$). Our present spectrum of Elara from 0.7-1.6 $\micron$ also compares closely with ground-based spectra of Himalia from \citet{Bhatt2017}. The band complexes in these objects indicate the presence of phyllosilicates and are discussed further in Section \ref{sec:himalia_fam}.

\begin{figure}[h!]
\begin{centering}
\includegraphics[width=0.65\textwidth]{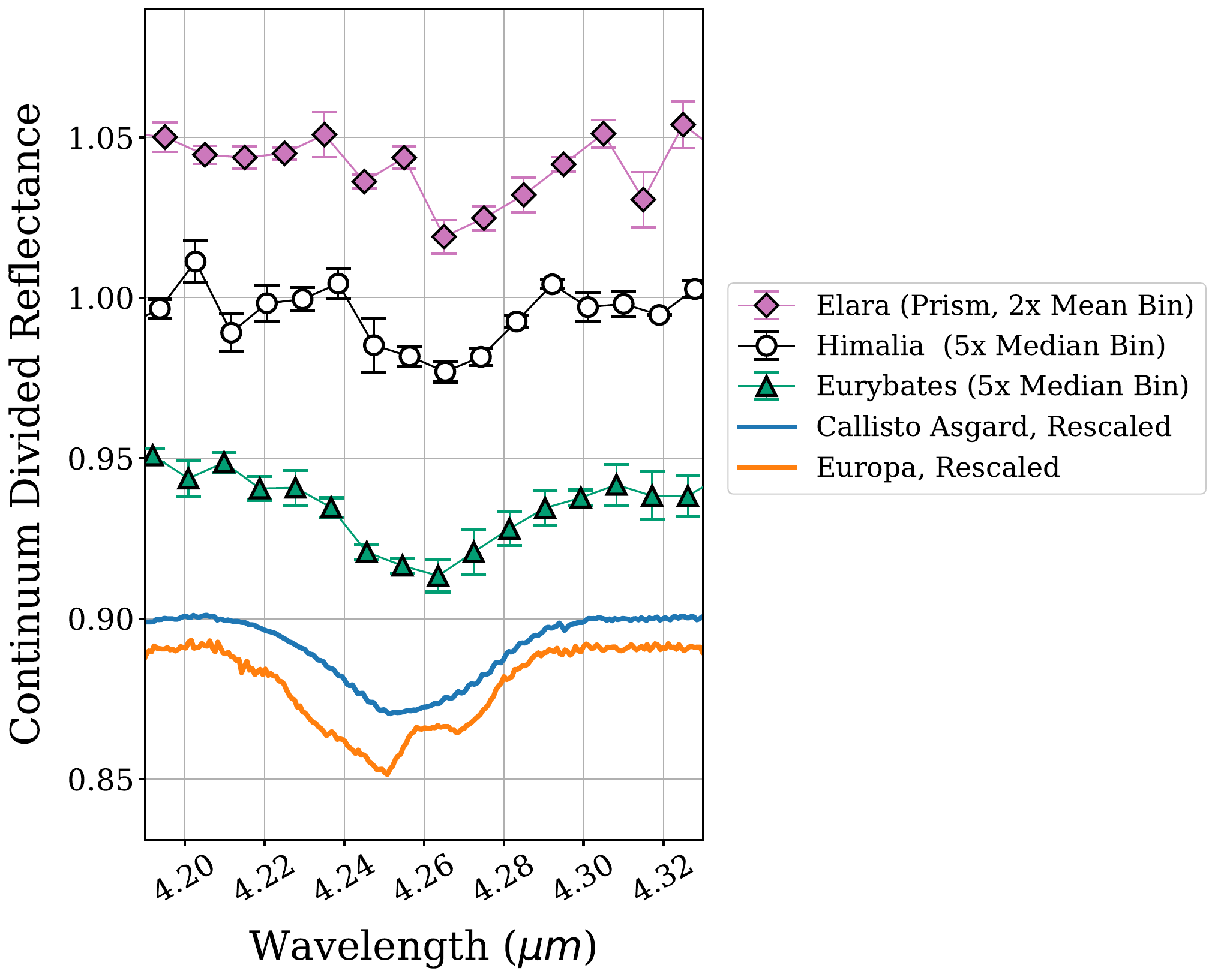}
\caption{Continuum-divided reflectance of the 4.15 to 4.40 $\micron$ region of Himalia and Elara.  They are compared with previous observations of Jovian Trojan Eurybates, which was reprocessed using the same reduction pipeline. The two irregular satellites show features near 4.27$\micron$ that are shifted $+\sim0.01\micron$ compared to Eurybates. The features on Himalia, Elara, and Eurybates are similar to but measurably distinct from the CO$_2$ bands observed on Europa and Callisto, which display minima at 4.25 and 4.27 $\micron$.}
\label{fig:425}
\end{centering}
\end{figure}

We also find a trace absorption at $4.27\micron$  on Himalia and Elara. This feature is similar to one discovered on the Jupiter Trojan Eurybates at $4.26\micron$ by \citet{Wong2024}. Figure \ref{fig:425} compares our observations of Himalia and Elara with the the \citet{Wong2024} observations of Eurybates, as processed via our reduction pipeline. We reprocess this spectrum only for methodological consistency between these datasets but do not note differences between reductions. The 4.26-4.27 $\micron$ features are similar to those caused by solid-state CO$_2$, which is volatile and unlikely to be stable as a crystalline ice on the relatively warm airless surfaces of Jovian satellites or Trojans \citep{Wong2024}. Since CO$_2$ is implausible as pure surface ice, it is likely complexed with more refractory components, as discussed for the origins of 4.25 and 4.27 $\micron$ features on the Galilean moons \cite[e.g.,][]{mccord1998,villanueva2023, cartwright2024, cartwright2025, trumbo2023distribution}. The discoveries of the $4.26-4.27\micron$ features on Eurybates, Himalia, and Elara place complexed CO$_2$ on several compositionally distinct small bodies near Jupiter.

The 3.0 $\micron$ absorber(s) for Carme, Sinope, Themisto, Pasiphae, and Lysithea are consistent with a variety of OH-bearing species, which we discuss in Sections \ref{sec:trojans} and \ref{ccs}. The absorption at 3.4 $\micron$ on Pasiphae, Ananke, Themisto, Sinope, and Carme is consistent with complex organics. All three objects in the Himalia family (Himalia, Elara, Lysithea) display a feature near 3.9 $\micron$ whose origins are unclear. Carbonates often present dual features near 3.9 and 3.4 $\micron$, which may be consistent with Elara but not Himalia nor Lysithea. We leave the causes of the Himalia family's 3.4 and 3.9 $\micron$ features as open questions in Table \ref{tab:BandProperties}. 

\section{Ammoniated Phyllosilicates in the Himalia Family} \label{sec:himalia_fam}

Himalia's complex spectrum from 2.6-3.8 $\micron$ strongly constrains its phyllosilicate composition. The sharp and deep feature at 2.7 $\micron$ can be reproduced with a variety of hydrated silicate materials from terrestrial and meteoritic inventories \citep[e.g.,][]{Takir2013,Takir2019,DeAngelis2021}. However, none of the hydrated silicates that match Himalia's 2.7 $\micron$ feature have the additional rounded minimum at $\sim$ 3.05 microns.

Some meteorite materials with significant amounts of adsorbed terrestrial water can reproduce the overall width of Himalia's band complex. However, we note that experimental studies of both meteorites and terrestrial phyllosilicates find that adsorbed water, and the spectral feature it causes, is easily removed when samples are treated under proper vacuum and desiccation procedures designed to mimic asteroid surface conditions \citep{Beck2010,Takir2013,Takir2019}. We therefore disfavor fits to Himalia that use phyllosilicates which are not treated to remove terrestrially adsorbed water.

\begin{figure}[h!]
\begin{centering}
\includegraphics[width=0.75\textwidth]{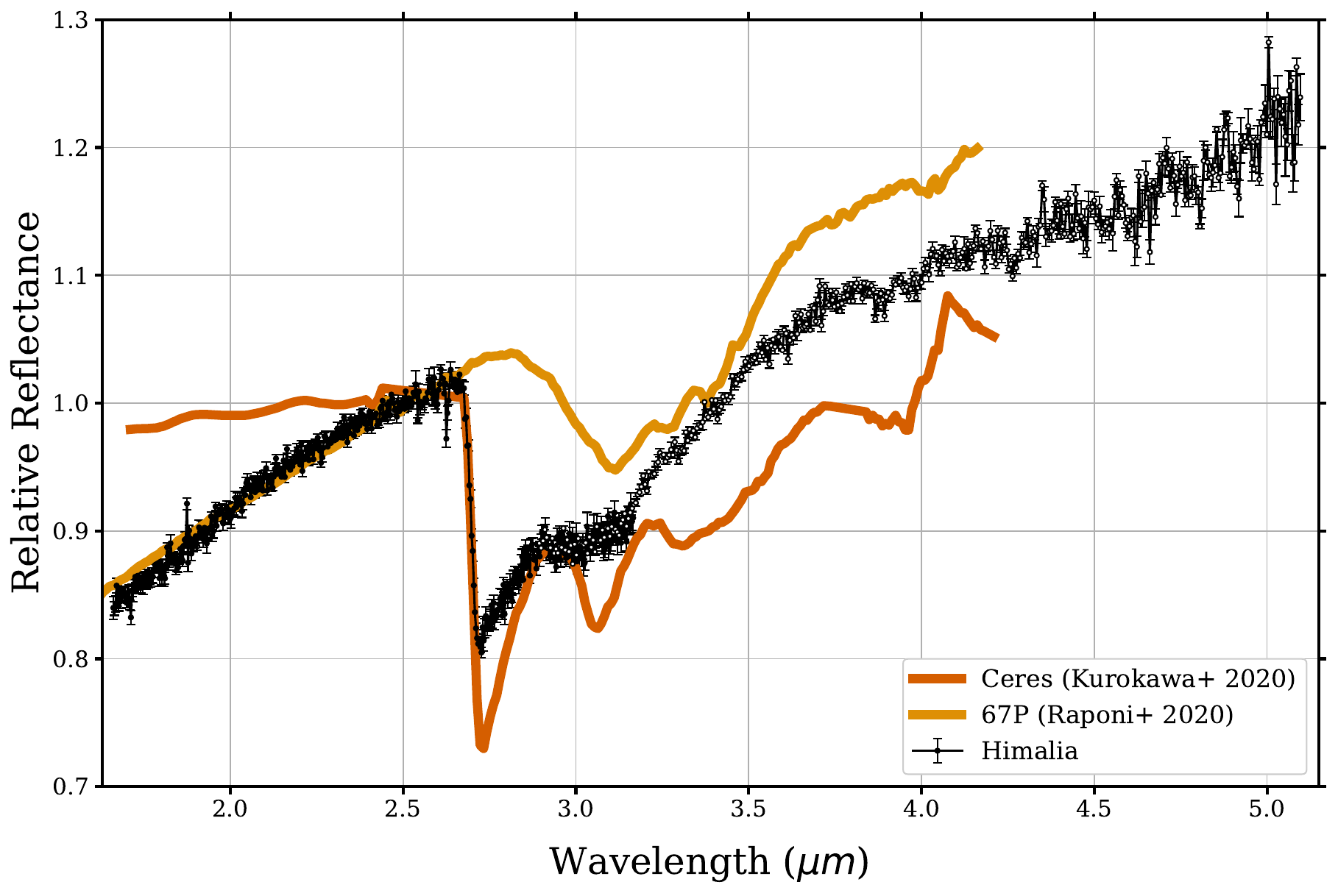}
\caption{Himalia compared with other small solar system bodies that show the 3.05 micron feature. While Himalia shows major spectral features similar to Ceres \citep{Kurokawa2020} at 2.7 and 3.05 microns, the differing relative strength of these bands suggest novel materials on Himalia which have not been catalogued on any other bodies at present. Faint features from 3.3-3.6 $\mu$m seen on 67P \citep{Raponi2020} are more similar to Himalia than corresponding deep and broad features  on Ceres.}
\label{fig:ammoniated_objects}
\end{centering}
\end{figure}

Himalia's combination of absorptions at 2.7 and 3.05 $\micron$ is reminiscent of Ceres \citep[][Figure \ref{fig:ammoniated_objects}]{Kurokawa2020}) and Hygiea \citep{Rivkin2025}, where the 3.05 $\micron$ feature is associated with the presence of NH$_4$ in the clay mineral structure \citep[e.g.,][]{King1992, rivkin2006,Ammannito2016,desanctis2016,desanctis2017}. Himalia, Ceres, and Hygiea have 2.7 $\micron$ bands with consistent centers \citep[2.72 $\micron$;][Table \ref{tab:BandProperties}]{deSanctis2015,Rivkin2025}. \citet{Pilorget2022} also detected this combination of 2.72 and 3.05 $\micron$ bands within the sample of asteroid Ryugu returned by \textit{Hayabusa2} mission. The Ryugu sample also contains some trace clasts with unaltered silicates \citep{Brunetto2023}, and the asteroid has been suggested to have formed near to the ice giants before its implantation into the main belt \citep{Hopp2022,Nesvorny2024}. Figure \ref{fig:ammoniated_objects} shows a comparison to 67P \citep{Raponi2020}, a presumably non-aqueously-altered object with absorptions between 3.0-3.5 $\micron$ attributed to ammonium-bearing salts \citep{Poch2020}. 

The 2.72 $\micron$ features on Ceres and Himalia are similar to some non-ammoniated meteoritic phyllosilicates. The center of the 2.7 $\micron$ feature caused by meteoritic phyllosilicates depends on the Mg/Fe content and therefore the degree of aqueous alteration (increased alteration corresponding to lower wavelength band centers). Initially anhydrous silicates become Fe-rich serpentines. Continued alteration increases phyllosilicate abundance and replaces Fe- with Mg to form antigorite and magnetite before ultimately forming clays like saponite \citep{Howard2011}. Heavily altered CI chondrites fit farther along this progression compared with typically less altered CM chondrites, though considerable variation exists in the 3 $\micron$ band shapes of these groups of meteorites \citep{Takir2019}. Ceres's $2.7 \micron$ feature is similar in strength and location to CI and heavily altered CM chondrites \citep{McSween2018}. Himalia's 2.7 $\micron$ band has the same center as Ceres's, suggesting both objects have phyllosilicates with similar Mg/Fe content. The band depth and shape of Himalia's 2.7 $\micron$ feature, however, may be more similar to the heavily altered CMs and C2-Ungrouped meteorites such as Tagish Lake and Essebi.


\begin{figure}[h!]
\includegraphics[width=0.49\textwidth]{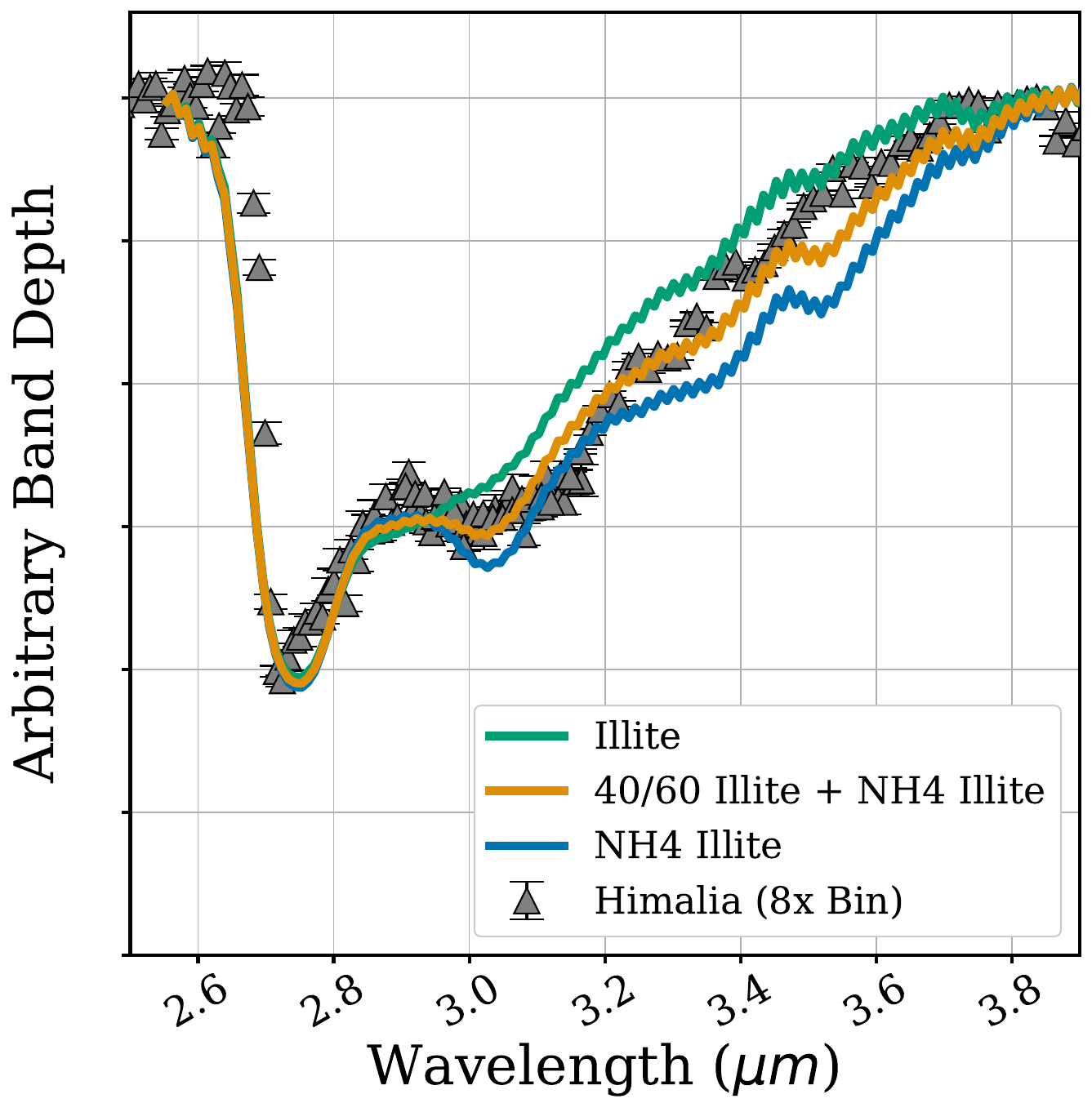}
\includegraphics[width=0.49\textwidth]{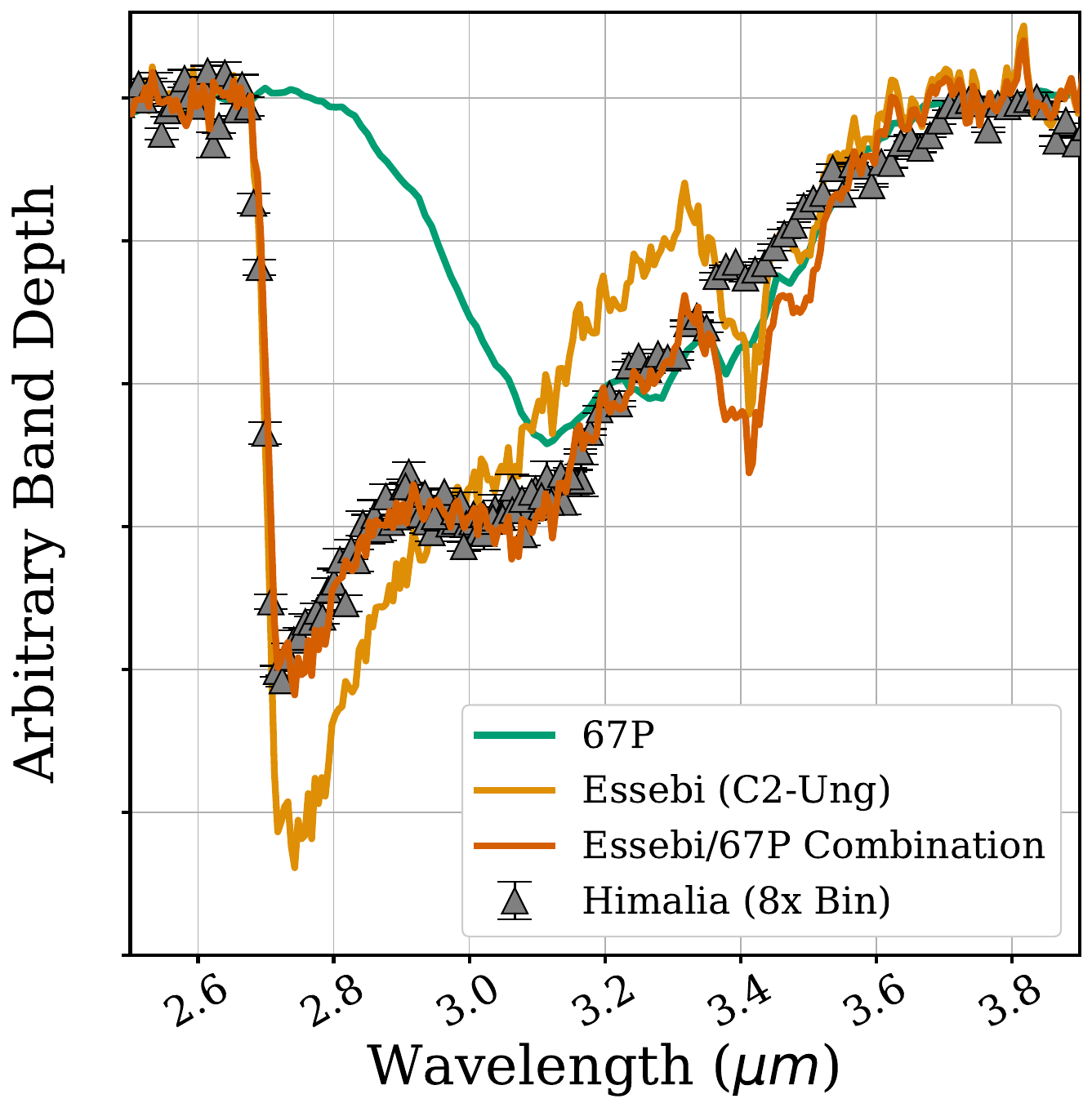}
\caption{Left: Comparison of Himalia’s continuum-divided 3.0 micron band complex to laboratory samples of ammoniated Illite from \citet{DeAngelis2021}. Note that absorption band depths are rescaled to approximate Himalia’s 2.7 micron feature, as pure laboratory samples have very large band depths. Right: Himalia’s 3.0 micron band complex is well matched by a combination of the phyllosilicate band found in aqueously altered C2-chondrite Essebi and the absorption band on 67P’s nucleus, attributed to ammonium-bearing salts. These materials do not form in the same geological settings, but they demonstrate that relatively simple mixtures of phyllosilicates with NH$_4$-bearing materials can provide a close spectral match to Himalia. The deep feature near 3.4 microns on Essebi is caused by aliphatic organics that are not present on Himalia, and is unrelated to phyllosilicate composition.}
\label{fig:himalia_comparisons}
\end{figure}

Himalia's 2.7 $\micron$ absorption feature shares some similarities with meteoritic materials, but the 3.05 $\micron$ feature has not been detected on any known meteorite. Instead, Himalia's combined 3 $\micron$ band complex is well-matched by laboratory samples of ammoniated phyllosilicates \citep{DeAngelis2021}. Figure \ref{fig:himalia_comparisons} shows that samples of illite, a clay mineral, can provide a reasonable fit to Himalia's spectrum from 2.6-2.9 microns. However, beyond 2.9 microns, illite's narrower absorption feature diverges from Himalia's wider band complex. When ammoniated in a laboratory setting, illite's absorption broadens and shows the key 3.05 micron absorption minimum. The presence of the 3.05 $\micron$ feature constrains the thermal history of Himalia's surface, as \citet{DeAngelis2021} found that the absorption feature only survives temperatures up to $\sim 600 K$. Such a temperature limit is consistent with the presence of Himalia's $0.7 \micron$ feature \citep{Jarvis2000,Vilas2024}, which disappears in CM2 materials heated $>400$ C (673 K) \citep{Hiroi1993} and has been linked to aqueous alteration processes that did not exceed 327 C (600K) \citep{Vilas1996}. We note that the ammonium content on Himalia is likely significantly lower than what is contained in the ammoniated-illite sample, as Himalia's spectrum is better described by a linear combination of untreated illite with the ammoniated sample (Figure \ref{fig:himalia_comparisons}).

Other, more complicated mixtures that include ammoniated materials can also fit the 2.7 and 3.05 $\mu$m bands. We do not consider the case of arbitrarily many absorbers (and thus arbitrary mixture complexity), but we do note that plausible matches are also provided by linear mixtures of non-ammoniated phyllosilicates with a separate ammonium-bearing material. The presence of such a mixture would imply that the phyllosilicates did not interact with NH-solutions during the parent body's aqueous processing, a geochemically incoherent situation if both materials were endogenic to the parent body. Yet this two-component example does provide a noteworthy spectral match, as illustrated in Figure \ref{fig:himalia_comparisons}. A combination of the hydrated C2-Ungrouped chondrite Essebi and an anhydrous ammonium-bearing material \citep[represented by 67P's nucleus][]{Kurokawa2020} provides a strong fit to the shape and location of Himalia's 3.05 $\micron$ feature. The Essebi/67P combination also reproduces the overall width and location of Himalia's 3 $\micron$ complex, but with two caveats: Essebi contains a strong organics feature at 3.4 $\micron$ and has a longer wavelength $2.7 \micron$ band minimum.

We conclude that hydrated silicates and ammonia-bearing materials (either as ammoniated phyllosilicates or as mixed components) provides the best fit to Himalia's band complex. We do not find any other single material that fits the overall shape of Himalia's entire band complex from 2.7-3.8 microns. For its simplicity as a single-absorber fit, we favor ammoniated phyllosilicates as the cause of the band complex. While the laboratory samples of illite provide a useful demonstration of the ammoniation process, meteorites display a variety of phyllosilicates in their matrices. Himalia's surface is likely to contain similarly complex phyllosilicate compositions.

\section{The Himalia Family's Compositional Heterogeneity}\label{himalia_fam}

\begin{figure}[h!]
\begin{centering}
\includegraphics[width=0.99\textwidth]{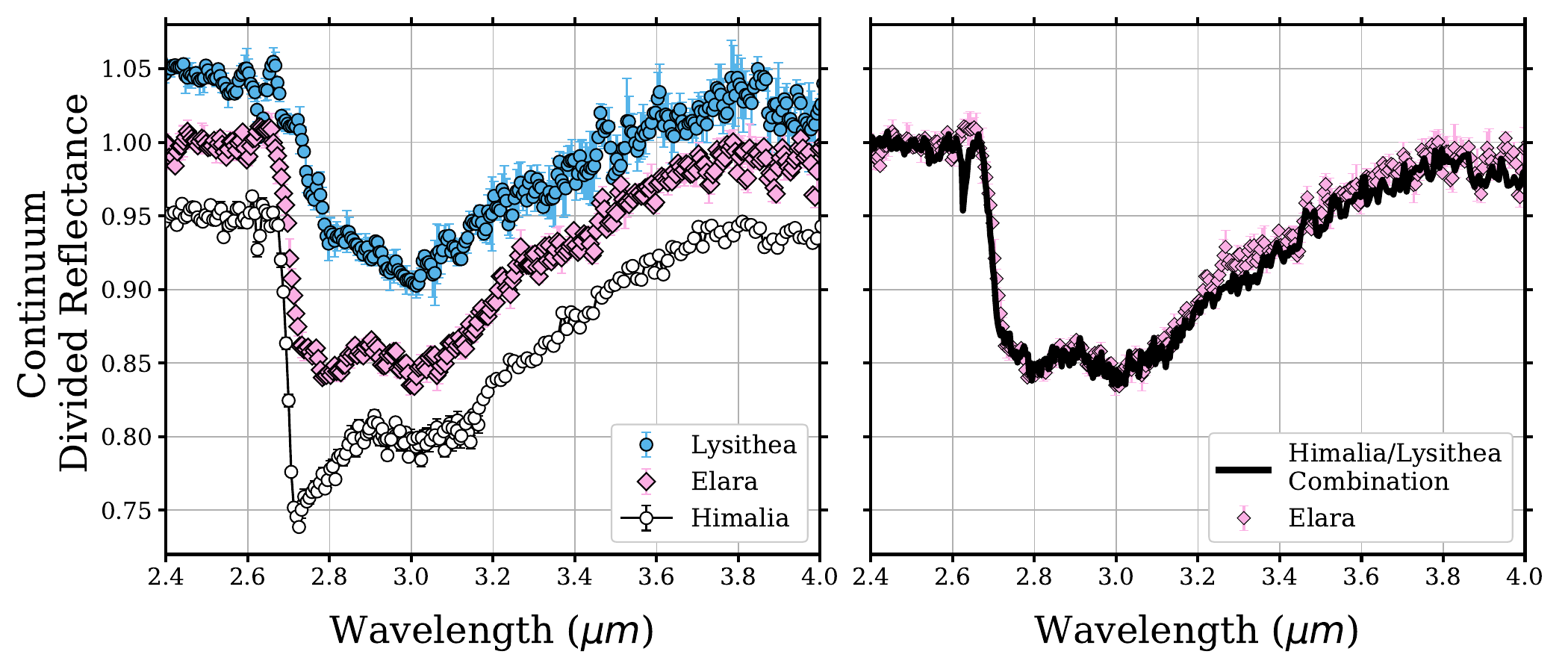}
\caption{The 3 $\micron$ regions of Himalia family objects (with shifts for Lysithea and Himalia of $\pm0.05$ for readability)}. Note that Lysithea displays a sharper band, overprinted by numerous subtle absorptions, while Elara and Himalia have more rounded features near 3 $\micron$. Bottom: We averaged Himalia's and Lysithea’s spectra into a new, combined spectrum that closely matches Elara, with the only exception being Elara’s lack of a 2.63 $\micron$ feature. The 3 $\micron$ band depth of the averaged Himalia/Lysithea absorption was multiplied by a factor of 1.07 to make comparison to the shape of Elara's spectral features easier.
\label{fig:elara_mix}
\end{centering}
\end{figure}

The Himalia family dominates the mass of Jupiter's irregular satellites by a factor of $\sim 10$ (assuming spherical volumes  with equal densities for all satellites). The Himalia family therefore stands as a major possible source for carbonaceous material within the Jovian system. But the bulk abundance of material within the Himalia family is unclear, as all three family members have different $3 \micron$ band complexes. If Himalia, Elara, and Lysithea formed from the same parent body, that would require that their parent experienced heterogeneous aqeuous alteration. Alternatively, their spectral variation may represent a mixture of materials from multiple parent bodies. Here, we explore whether the observed spectral properties of Himalia, Elara, and Lysithea favor a single-parent or multi-parent origin.

A comparison of the Himalia family members' continuum-removed $3 \micron$ region is shown in Figure \ref{fig:elara_mix}. Lysithea presents a single-minimum band, while Himalia and Elara display composite features with minima near 2.7 and 3.0 $\micron$. The Himalia family's band properties do not just vary in depth: Himalia has a rounded $3.0 \micron$ band that is centered $\sim0.05 \micron$ longer than Lysithea's sharper feature (see Table \ref{tab:BandProperties}. Elara's absorption feature is similar to a simple (50/50) mean of Himalia and Lysithea's band complexes (Figure \ref{fig:elara_mix}, Right). The Himalia/Lysithea blend describes both Elara's 2.7 $\micron$ feature and Elara's intermediate band shape/center from 2.9-3.2 $\micron$. Elara's surface appears to contain both Lysithea's unknown 3.0$\micron$ absorber and the ammoniated phyllosilicates present on Himalia.

While Elara's $3 \micron$ band complex is intermediate between Himalia and Lysithea, its overall spectrum is not. Elara's spectral slope matches Himalia, while Lysithea diverges from 0.7-1.5 $\micron$ compared with both Elara (Figure \ref{fig:alldata}) and Himalia \citep{Bhatt2017}.  The presence of trace absorption features add further complexity: Himalia and Elara display 4.27 $\micron$ features; Himalia, Elara, and Pasiphae all display 3.9 $\micron$ features; Himalia and Lysithea display 2.63 $\micron$ features. The Himalia family members present a complex compositional picture whose only clear spectral trend is the correlation between the presence of ammoniated materials and satellite size.

Elara is not a simple mixture of Lysithean and Himalian materials, but it does show that there is some relationship between their 3 $\micron$ absorbers. We suggest that Elara's combination of Himalia-like colors and its blend of Himalian/Lysithean 3$\micron$ bands is best understood as a distinct processing state of material from the Himalia parent body. This interpretation favors the hypothesis that all three objects were primarily formed from the same parent body, with the three objects representing distinct aqueous alteration states. The origins of trace absorption bands remain unclear, but could relate to the specifics of parent body alteration processes. 

Regardless of specific origins, the Himalia family contains at least 2 distinct types of OH-bearing materials (Himalia-like and Lysithea-Like) that do not correlate with optical or near-infrared color. If Himalia, Elara, and Lysithea all reflect material from the same parent body, our observed progression of hydration features agrees with the observations and hypothesis of \citet{Vilas2024}, who proposed that these bodies represent the parent body's core, mantle, and crust material, respectively.

Future searches for surface heterogeneity and measurements of smaller family members are necessary to constrain family variability. It is yet unknown if there are any consistent spectral types within the Himalia family or if there is instead a continuous array of compositions between separate endmembers. Constraining compositional trends would be useful to determine how aqueous alteration occurred on the Himalia parent body. Alteration may have occurred as a function of depth on some irregular satellite parent bodies \citep[e.g.,][]{Vilas2024,Sharkey2023} or perhaps in more localized pockets \citep[e.g., like the CV3 Mokoia parent body,][]{Tomeoka2011}.

We propose that the Himalia family parent body was similar to Ceres in terms of containing water, organics, and NH-bearing materials. Furthermore, the location of Himalia's 2.7 $\micron$ feature is in close agreement with the same feature on Ceres, suggesting both objects experienced similar aqueous alteration processes. Yet Himalia and Ceres differ in the shape and depth of the 3.05 $\micron$ region. It is possible that the spectral distinctions reflect differences in initial abundances of NH-compounds, but Ceres and Himalia have also likely experienced divergent geological processing. Discerning whether or not Himalia and Ceres could have formed from the same primordial reservoirs constrains both objects' migratory histories and could position the smaller Himalia family members as possible samples of less-processed Ceres-like objects.

The ultimate origins of Himalia, and by extension NH in the Jovian system, are still unclear. Beyond the comparisons to Ceres and other inner-solar-system asteroids, we also note similarities between Himalia and the Jovian inner satellite Amalthea from $2.9-4.0 \micron$ \citep{Takato2004}. Amalthea's reflectance spectrum has not been characterized between $2.5-2.9 \micron$, limiting the constraining power of this comparison considerably. However, the similarities in possible shape and location of Amalthea's absorption band to Himalia's are intriguing. Amalthea has a low-eccentricity and low-inclination orbit interior to the Galilean satellites, and it is thought to represent material that accreted close to Jupiter during the planet's formation \citep{Greenberg2010}. Understanding if Himalia includes any Amalthea-like materials, which may have accreted prior to Jupiter's hypothesized migration, would inform whether or not Himalia itself formed near to Jupiter prior to its capture. Given the potential implications for Himalia's provenance, as well as the possible effects of Himalian dust on Galilean surfaces, future spectroscopic assessment of Amalthea and the other small inner satellites is crucial to understand the Jovian satellite system as a whole.

\section{Implications for Nitrogen Delivery to the Galilean moons}

NH$_4$-bearing species on Himalia and Elara represent an important source of nitrogen in the Jovian system, which could be delivered to the Galilean moons in dust grains migrating inward on slowly decaying orbits due to Poynting-Robertson drag \citep{burns1979}. Prior ground and space-based observations have suggested that Callisto's 4.57 $\micron$ feature may result from CN-bearing (or CS-bearing) compounds delivered in dust from the irregular satellites \citep{cartwright2020, cartwright2024}. 

However, we detect no evidence for a 4.6 $\micron$ absorption feature, nor any reliable evidence for S-bearing species, in the irregular satellite spectra reported here. Instead, NH$_4$-bearing compounds delivered in dust grains may provide an important source of nitrogen that mixes with H$_2$O ice and native (or delivered) C-rich material in Callisto's regolith. This mixture of N, C, H, and O is then irradiated by charged particles trapped in Jupiter's magnetosphere (including S$^+$ ions), forming new radiolytic products, possibly including OCN$^-$ and long-chain, refractory organic residues that express C-H and C-N stretching modes \citep[e.g.,][]{accolla2018}. 

Delivery and subsequent irradiation of NH$_4$-bearing material could contribute to the subtle absorption features detected between 3.2 and 3.7 $\micron$ and Callisto's broad 4.57 $\micron$ band. Recent dynamical modeling of dust grain trajectories indicates that dust from Jupiter's retrograde irregular satellites should primarily strike Callisto's leading hemisphere, whereas dust from prograde Himalia might collide with all four of the Galilean moons (Figure 8 in \citealt{chen2024}). Although NH$_4$-rich dust from Himalia could remain on the surface of Callisto and darker regions of Ganymede over long geologic timescales, this NH$_4$-rich content might be incorporated into Europa's interior via available geologic conduits, such as chaos-dominated terrains in Tara Regio and elsewhere, over much shorter timescales \citep[e.g.,][]{hand2009,hesse2022}, potentially providing an exogenic source of nitrogen to Europa's subsurface ocean \citep{cartwright2024}, a key element for life as we understand it.

\section{Trojan-like Surfaces Exist Amongst Fragmented Irregular Satellites}\label{sec:trojans}

\begin{figure}[h!]
\begin{centering}
\includegraphics[width=0.95\textwidth]{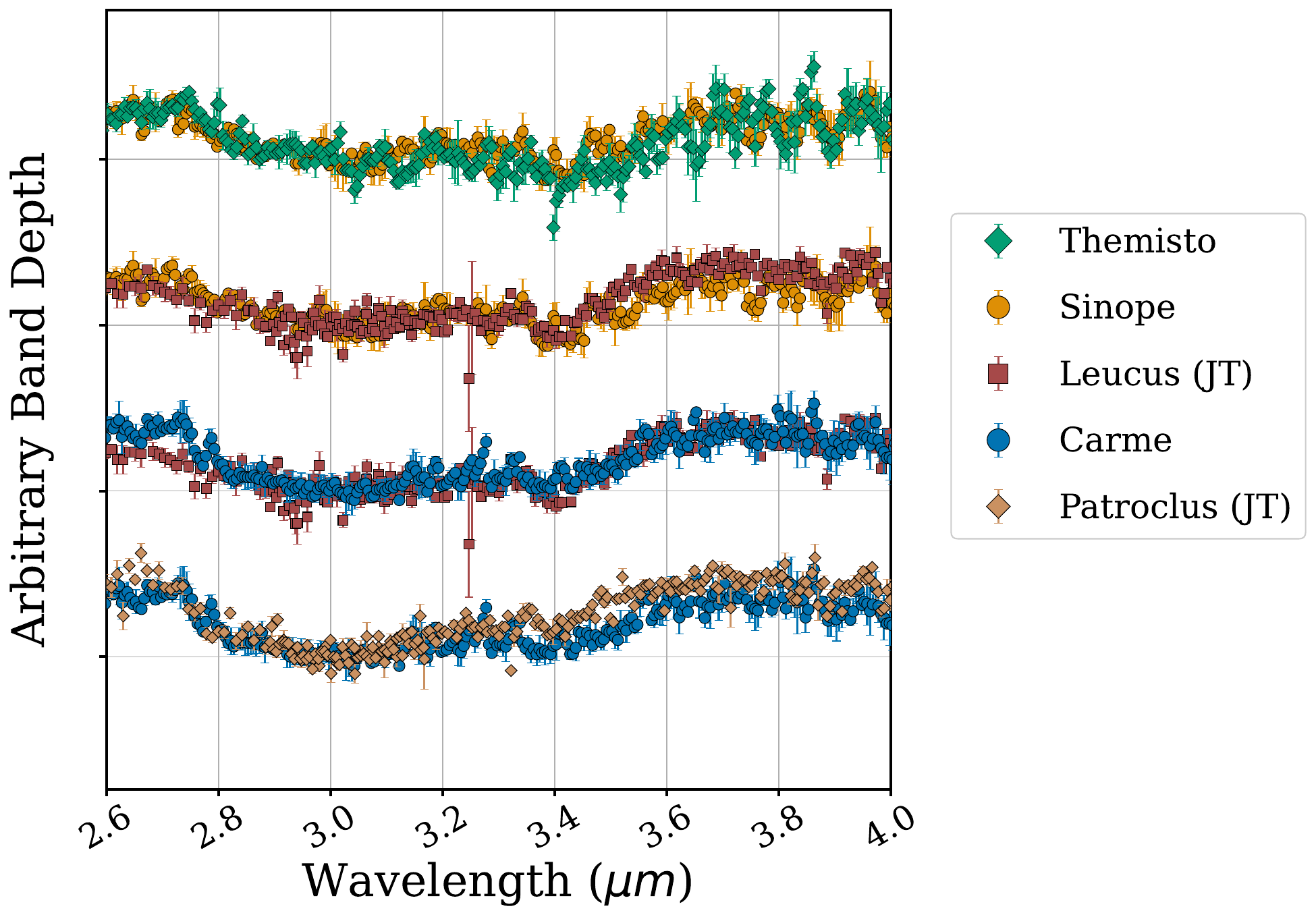}
\caption{Comparisons of $3 \micron$ band complexes between irregular satellites Themisto, Sinope, and Carme to the Jovian Trojans Leucus and Patroclus. Band depths for all objects were normalized at $3.0 \micron$ to aid comparisons of the relative strengths and shapes of the 3.0 and 3.4 $\micron$ bands for each object. Themisto, Sinope, and Leucus all belong to the ``red'' color taxon and display similar 3.0 and 3.4 $\micron$ absorption features. Carme's band properties are intermediate between ``red'' and ``less-red'' groups, with a deeper 3 $\micron$ feature that compares closer in shape and band center to the ``less-red'' Trojan Patroclus, but a deeper 3.4 $\micron$ feature from aliphatic organics that compares closer to the ``red'' objects.} Trojan spectra are reprocessed for methodological consistency and were originally reported and analyzed in \citet{Wong2024}.
\label{fig:red_trojan_comp}
\end{centering}
\end{figure}

Previous assessments of the spectrally ``red'' irregular satellites have relied on color analogies to Jovian Trojan asteroids without specific constraints from absorption features. VNIR measurements \citep{Sharkey2023,Grav2004a} found that irregular satellites outside of the Himalia family map within the color taxonomy of Jovian Trojans. Figure \ref{fig:red_trojan_comp} compares the irregular satellites Themisto, Sinope and Carme with the Jovian Trojans Leucus and Patroclus observed by \citet{Wong2024}. These JWST observations demonstrate that the reddest irregular satellites indeed share common absorption features at 3.0 and 3.4 $\micron$ with their Trojan counterparts.

Sinope and Themisto both have 3.0 and 3.4 $\micron$ bands that correspond directly to those seen amongst the red Trojans (see Themisto/Sinope and Sinope/Leucus comparisons in Fig. \ref{fig:red_trojan_comp}). However, Carme, the largest member of its own collisional family, displays a surprising combination of band properties. Carme appears to have a different ratio of organics to hydroxylated minerals than seen in either the redder or less-red Trojan groups, with the organics apparently typical of the redder objects and the hydroxylated minerals typical of the less-red objects. \citet{Wong2024} found that the less-red objects Patroclus and Eurybates display weaker 3.4 $\micron$ organics bands and shorter-wavelength 3.0$\micron$ bands as compared with the red objects Orus and Leucus. VNIR data discriminates Carme into the ``red" taxon and differentiates it from the ``less-red" Patroclus \citep{Emery2003,Sharkey2019}. Carme further matches the red Trojan group by having a stronger 3.4 $\micron$ band. But Carme's 3.0 $\micron$ band clearly differs from the long-wavelength absorption seen on red Trojans, and instead has a shape and central wavelength matched by Patroclus (Fig. \ref{fig:red_trojan_comp}).

Carme's $3.0 \micron$ band location does not correspond to those seen on any similarly colored objects. Carme's properties are interesting to consider in the overall context of Trojan color groups. Carme's spectral properties suggest that spectral slopes of Trojans are controlled by organic abundance and less controlled by the 3-micron absorber. Alternatively, Carme's band properties themselves could have been altered via collisions, perhaps from implanting exogenous material or from revealing interior materials with differing 3.0 $\micron$ band properties. Determining Carme's history at this point is difficult, however, without additional context from its other family members' band properties, as well as a broader understanding of the variation within Trojan $3.0 \micron$ band properties.

The presence of Trojan-like 3.0 and 3.4 $\micron$ absorption bands on small Jovian irregular satellites provides new context for the chemical pathways that could have produced them. The irregular satellites' extensive collisional processing \citep{Bottke2010} took place after their capture about Jupiter. That the irregular satellites fragmented post-capture requires that their current surfaces were created and evolved within the circumjovian environment. This may differ from similarly sized Trojans, as Trojans larger than  $\sim 10\ km$likely experienced more collisions in the primordial Kuiper belt than in their current locations \citep{Nesvorny2018,Marschall2022,Bottke2023}. If the 3.0 and 3.4 $\micron$ absorbers were present on irregular satellite parent bodies prior to their capture, then those absorbers must have been present in high enough abundances to survive collisional fragmentation and be incorporated into the surfaces of the present day population. Alternatively, the 3.0 and 3.4 $\micron$ absorbers could be produced in-situ on satellite surfaces via irradiation processing \citep[as discussed by][]{Wong2024}. In-situ surface formation of the 3.0 and 3.4 $\micron$ absorbers would require that they be produced by materials that are stable within the Jovian thermal environment.

\section{What causes Less-Red OH Absorptions?} \label{ccs}

\begin{figure}[h!]
\begin{centering}
\includegraphics[width=0.65\textwidth]{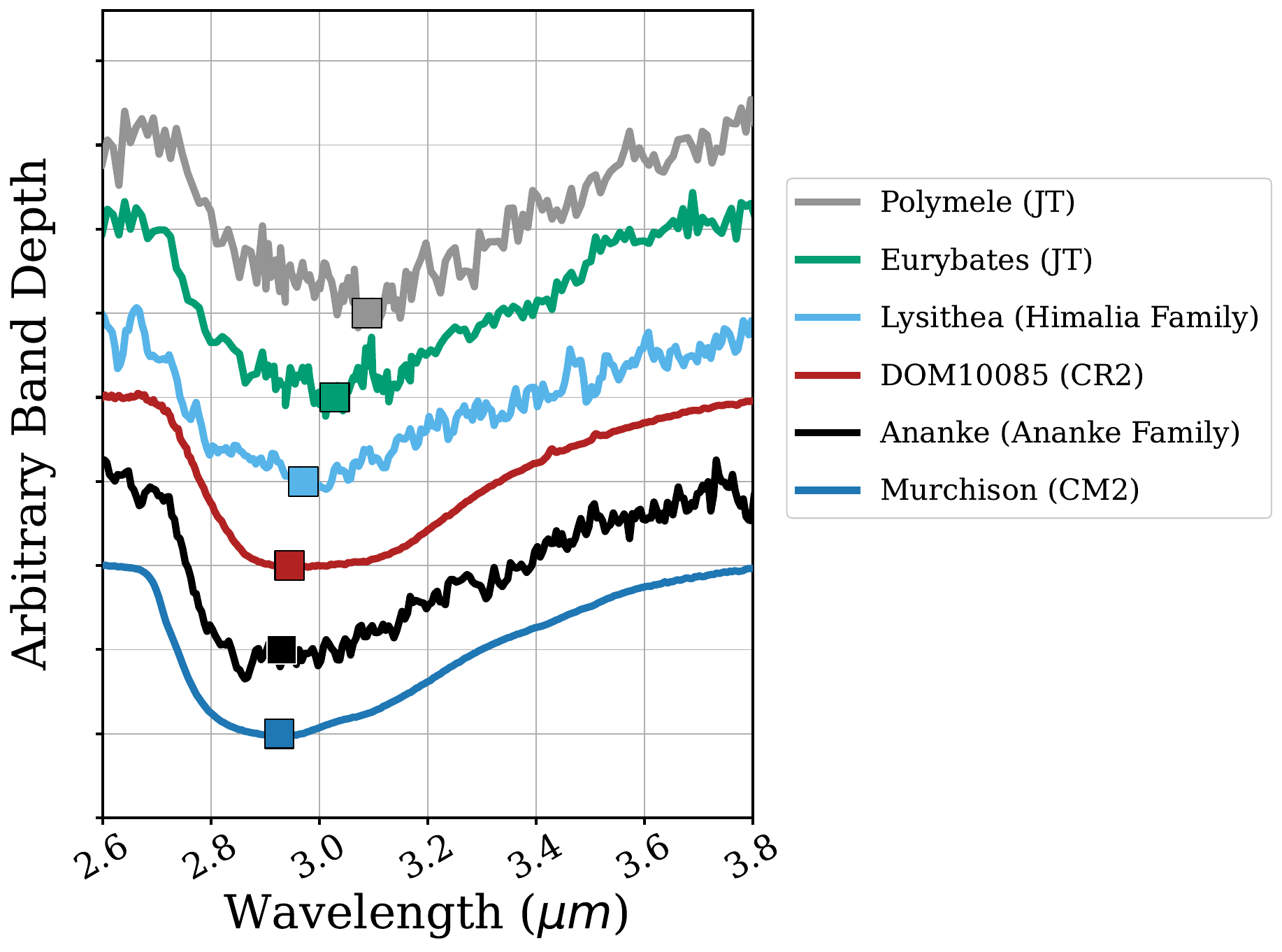}
\caption{The 3.0 $\mu$m band shapes for: irregular satellites Lysithea and Ananke; meteorites Murchison and DOM 10085; Jupiter Trojans Eurybates and Polymele. Objects are sorted according to the central wavelength ($\lambda_c$) of their features, which is marked by a square on each curve. Ananke's feature is positioned in between known petrologic type-2 meteorites that have experienced aqueous alteration. Lysithea is centered intermediate between Eurybates and Ananke. For ease of band shape comparisons, all bands are rescaled to the same arbitrary depth.}
\label{fig:gray_trojan_comp}
\end{centering}
\end{figure}

Ananke, Lysithea, and Pasiphae present $3 \micron$ absorptions with central wavelengths $\lambda_c \sim2.95 \micron$. These absorptions are broadly associated with OH-bearing compounds, but we do not find any specific materials that match them directly. Comparisons to other OH-absorptions on small bodies and meteorites show that the mystery of the $\sim 2.95 \micron$ absorptions may be related to similar questions on a wide range of solar system surfaces. Specifically, the less-red irregular satellites Ananke, Lysithea, and Pasiphae fill an apparent gap between known Trojans ($\lambda_c > 3.0\micron$) and any currently identified groups of main belt asteroids \citep[e.g.,][who grouped objects between those with $\lambda_c > 3.0\micron$ and those with estimated $\lambda_c \sim 2.7-2.9 \micron$ from ground-based observations]{Rivkin2022}. 

Figure \ref{fig:gray_trojan_comp} places these less-red irregular satellite absorptions in comparison with two aqueously altered meteorites \citep{Takir2019,Yu2024} and two Trojans \citep{Wong2024}, sorted by their $\lambda_c$. The comparison meteorites were chosen as they represent some of the longest-wavelength absorptions yet observed on aqueously altered primitive materials, which typically cluster near 2.7-2.8 $\micron$ \citep{Takir2019}. Eurybates and Polymele are chosen as they are representative of the two types of absorptions seen amongst less-red Trojans \citep{Wong2024,Brown2025}.

The relationships between the spectra in Figure \ref{fig:gray_trojan_comp} have broad implications for ice-poor surfaces throughout the outer solar system. Within the Jovian system, the prograde Lysithea (Himalia Family) is highly similar to the retrograde satellite Pasiphae (Pasiphae Family), as shown in Fig. \ref{fig:ly_pas}. The Lysithea/Pasiphae 3 $\micron$ absorber exists within two separate families with dramatically different compositional contexts, suggesting that this material existed on multiple irregular satellite parent bodies. Outside of the Jovian satellite system, Eurybates and Polymele have absorption bands centered $\lambda_c>3.0 \micron$ similar to the ice-poor Saturnian irregular satellites Albiorix and Siarnaq \citep{Belyakov2025}. These Saturnian satellites were themselves noted by \citet{Belyakov2025} to be similar to the ice-poor Neptune Trojan 2006 RJ103 \citep{Markwardt2023}.

Ananke's $3\micron$ band is centered between the petrologic type-2 chondrites CM2 Murchison \citep{Yu2024} and CR2 DOM 10085 \citep{Takir2019}. The broad, rounded feature on Murchison is attributed to the presence of structural OH in phyllosilicates. Ananke's 3 micron band appears similar to this Murchison-like absorption, and we conclude that it is consistent with these known C2 phyllosilicate assemblages. \citet{Vilas2024} report a 0.7 $\micron$ absorption band on one of two collected observations of Ananke caused by phyllosilicates.

Understanding Lysithea's composition is challenging, as its $3\micron$ band center is intermediate between the shorter-wavelength meteorite absorptions and the longer-wavelength Trojan absorptions (see centers indicated in Fig. \ref{fig:gray_trojan_comp}). Previous observations have suggested that Lysithea (and the spectrally similar Pasiphae) contain phyllosilicates \citep{Vilas2024}, and Lysithea appears related to the otherwise aqueously altered Himalia family. However, Lysithea's $3 \micron$ band center is not clearly reconcilable with alteration products from the meteorite record. Some very weakly altered meteorites (petrologic type 3) have minor amounts of phyllosilicates and display absorptions centered $\lambda_c > 3\micron$ \citep{Takir2013,Takir2019}. However, these meteorite absorptions can be sensitive to major distortions by terrestrial contamination, particularly from Fe-oxyhydroxides. Lysithea's absorption therefore cannot be entirely understood by analogy to C-chondrite like phyllosilicates. Instead, it is possible that Lysithea contains a mixture of Trojan-like OH compounds and aqueously altered phyllosilicates, which combine to create the shifted band. Such a scenario remains highly speculative, however, as it does not lead to any directly testable spectral parameters at present.

Regardless of the bands' origins, Lysithea ($\lambda_c \sim2.97\micron$), Eurybates ($\lambda_c \sim3.03\micron$), and Polymele ($\lambda_c \sim3.08\micron$) present a possible progression of band centers. It is not clear how closely associated these objects' compositions could be, but we note that the scale of variation between the two Trojans (Eurybates and Polymele) are approximately the same ($\sim0.05\micron$ shifts) as the differences between Eurybates and the less-red satellites Lysithea/Pasiphae. Observationally, we do not yet understand if the less-red objects generally present a continuum of 3 $\micron$ bands or if any clustering of spectral parameters exist. Without a broader understanding of the $3 \micron$ band center variability amongst the Jovian Trojans, it is difficult to further assess the Lysithea/Eurybates band differences. Future characterization of the Eurybates collisional family \citep{Wong2025} will provide a key opportunity to understand if Lysithea-like absorptions exist amongst the Trojans, or if this band is unique to the Jovian satellites.

Ultimately, the origins of $3 \micron$ absorptions on less-red Jupiter Trojan and irregular satellite surfaces remain ambiguous. Previously, \citet{Wong2024} noted that the 3 $\micron$ bands of Trojans are generally similar to some spectra of radiolytically processed C- and N-bearing ices, hypothesizing that this evolutionary pathway could fit their detections of Trojan OH absorptions and their nondetections of water ice. The application of the ice irradiation hypothesis to the irregular satellites is more complex, however, as this scenario would require that the OH-bearing weathering product(s) survived collisional resurfacing or were reproduced in the Jovian environment, as discussed in Section \ref{sec:trojans}.

\section{Conclusions}

\begin{figure}[h!]
\begin{centering}
\includegraphics[width=0.75\textwidth]{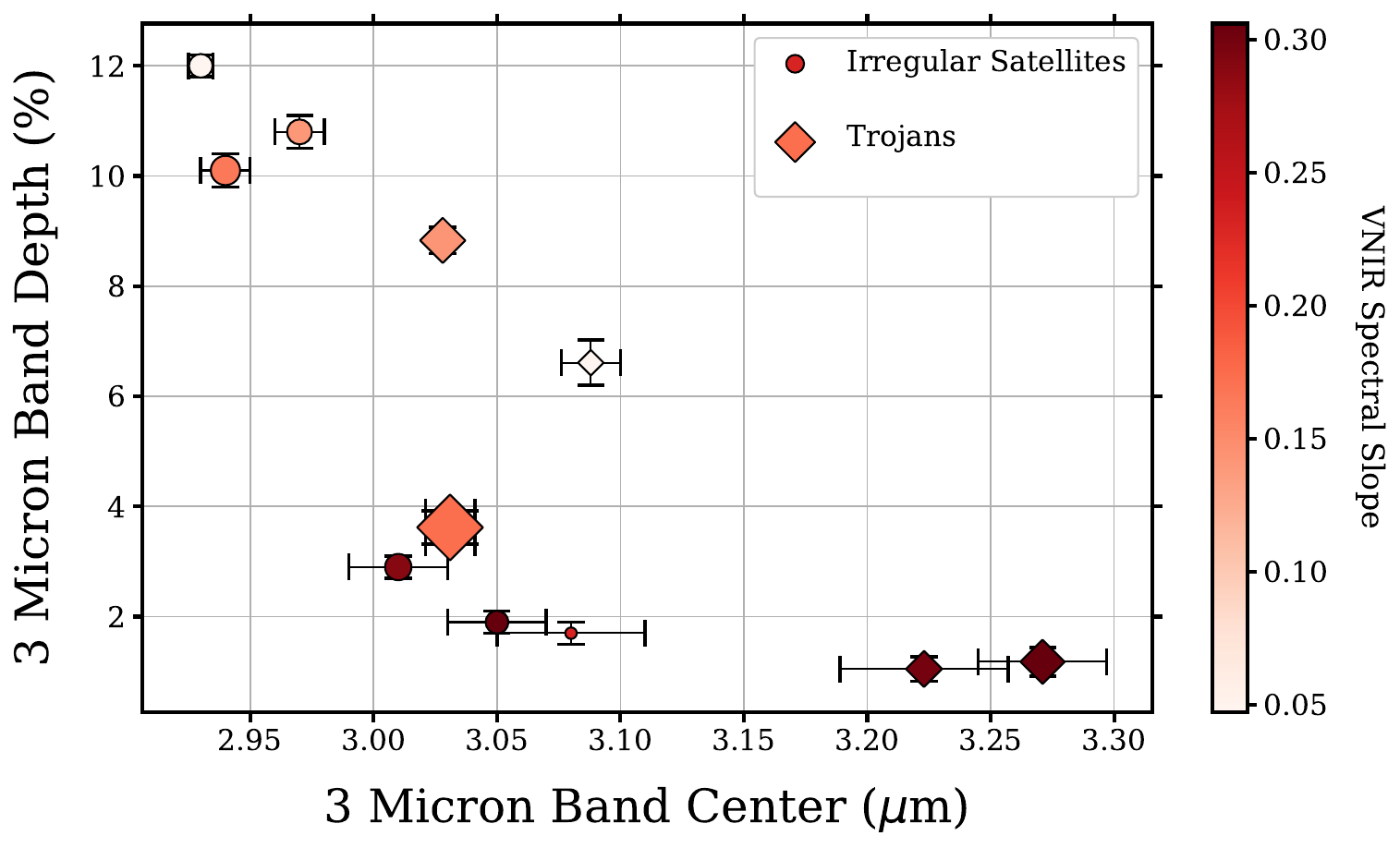}
\caption{The 3 $\micron$ band centers vs. band depths of Jovian Trojan and irregular satellites with available band analysis parameters, from this work and \citep{Wong2024}. Himalia and Elara are excluded from this analysis due to their overlapping features at 2.7 and 3.0 microns, which render a single band center ambiguous. Marker color corresponds to the object's spectral slope from 0.7-2.4 $\micron$ collected this work or from the literature \citep{Wong2024,Sharkey2019}. Marker size refers to the diameters as collected in Table \ref{tab:circumstances}.}
\label{fig:bandpars}
\end{centering}
\end{figure}

We report 0.7-5.2 $\micron$ reflectance spectra of eight Jovian irregular satellites collected with JWST's NIRSpec instrument. Despite similar hypotheses for their origins, the irregular satellites are more compositionally diverse than the currently known Jovian Trojans. At least three surface types are present amongst the irregular satellites: Ceres-like surfaces, red-Trojan-like surfaces, and less-red surfaces which are intermediate between Trojans and hydrated asteroids/carbonaceous chondrites.

The two largest irregular satellites have major features similar to those of Ceres-like main-belt asteroids. Himalia and Elara ($\sim$ 140 km and $\sim$ 80 km, respectively, both in the prograde Himalia family) have bands at 2.7 and 3.05 $\micron$ demonstrating the presence of ammoniated phyllosilicates. Elara's major band properties are consistent with a simple mean of its family members Himalia and Lysithea. Himalia and Elara also have a minor feature at 4.27 $\micron$, similar to the feature discovered on the Trojan Eurybates and associated with complexed CO$_2$ \citep{Wong2024}.

We propose that the Himalia parent body formed with materials similar to Ceres-like ammonium-bearing asteroids. It is not clear how closely linked Himalia and Ceres may be without further understanding the distribution of ammoniated materials in small body populations. We note Himalia is spectrally similar to ground-based observations of the inner satellite Amalthea \citep{Takato2004}, which may represent the primitive materials that accreted near Jupiter prior to its migration. Further compositional characterization of the inner satellites is critical to assess whether Jupiter captured and retained any irregular satellites from zones near to the planet's formation region.

The observed spectral variation in the Himalia group, including Elara's blend of Himalia-like and Lysithea-like material, could relate to a stratified mutual parent body, similar to the hypothesis of \citet{Vilas2024}. We suggest that the complex hydration states observed within the Himalia family are more consistent with a heterogenous parent body and disfavor separate parent body origins for the 2.7 and 3.05 $\micron$ absorbers.

In addition to the ammoniated phyllosilicates in the Himalia family, at least two other distinct types of OH-bearing materials are ubiquitous among both prograde and retrograde irregular satellites. Figure \ref{fig:bandpars} summarizes the band properties of these irregular satellites (excluding Himalia and Elara) in comparison to similar measurements of Jovian Trojans \citep{Wong2024}. The less-red irregular satellites (those with smaller VNIR slopes, plotted with lighter shading in Figure \ref{fig:bandpars}) have shorter wavelength 3 $\micron$ bands compared to their counterpart Trojans. The red irregular satellites (higher VNIR slopes and darker shading in Figure \ref{fig:bandpars}) also display shorter wavelength absorptions compared to their Trojan equivalents, although their bands are less deep and may therefore be more prone to systematic uncertainties in their band centers. We do not attempt to draw any specific trends in the band parameters shown in Figure \ref{fig:bandpars}, but note the wide range of colors, band depths, and band centers that are present amongst these materials.

The ``red" group of satellites (Carme, Sinope, Themisto) display one notable outlier compared with ``red'' Trojans. Carme has a similar $3 \micron$ band to the less-red Trojan Patroclus, but a deeper 3.4 $\micron$ organics band more similar to those seen amongst the red Trojans as discussed in Section \ref{sec:trojans}. The prograde satellite Themisto and the retrograde satellite Sinope have 3.0 and 3.4 $\micron$ bands similar to the red Trojan Leucus. The colors of red objects appear to correlate with the depth of the 3.4 $\micron$ feature. We do not propose specific compositions for the ``red'' irregular satellites, but we note that they are the products of intense collisional evolution after their capture about Jupiter \citep{Bottke2010}. Therefore the ``red'' 3.0 and 3.4 $\micron$ absorbers must either form within the circumjovian environment or represent material that survives major resurfacing events.

The less-red irregular satellites defy simple classification, with bands that are similar to the less-red Trojans but at measurably shorter wavelengths. The retrograde satellite Ananke is similar to some hydrated chondrites, suggesting that aqueous alteration is present within the retrograde satellite population and is not limited to the prograde Himalia family. Pasiphae (Pasiphae Family) has a 3 $\micron$ absorption feature that is similar to Lysithea (Himalia Family). The origins of this absorber are unclear, and its presence amongst two separate families complicates a clear association with a single surface type. We speculate that this group may contain Trojan-like materials that have experienced some weak aqueous alteration to modify their OH absorption bands.

We suggest two possible methods to explain the overall variation of the irregular satellites and their distinctions compared to the Jovian Trojans. The first possibility is that the Himalia and/or Ananke families represent separate compositional lineages that were not captured into the Jovian Trojans. In this case, the capture of the Himalia and Ananke parents would be unrelated to the scattering of the primordial Kuiper belt (the hypothesized origins of the Trojans). Instead, Himalia and Ananke could represent captures from an earlier epoch of Jupiter's history. This hypothesis can be tested by understanding the compositions of Jupiter's inner satellites, whose current surfaces may shed further light on the materials that accreted near to Jupiter's formation region. 

Alternatively, for Himalia and Ananke to be related to Trojan compositions, they could  represent the unseen cores of Trojan-like objects. A Trojan/Himalia relationship may seem implausible due to the lack of any signs of ammoniation amongst the Jovian Trojans. Similarly, Ananke's lower wavelength 3 $\micron$ band is unlike any that has been observed on a Trojan. However, the possible presence of Trojan-like material (Lysithea) within the Himalia family suggests that some relationship between Himalia and Trojan-like materials may exist. Connecting either or Himalia to Trojan-like compositions requires that Lysithea-like OH absorptions are directly related to those of the less-red Trojans, despite their different wavelength centers.

Either origin scenario for Himalia and Ananke remains speculative without understanding the nature of $3 \micron$ variation among any other asteroid collisional families. Future tests for 3 $\micron$ variability among collisionally evolved Trojans \citep{Wong2025} and main-belt asteroids \citep{Takir2024} will be useful to explore compositional stratification amongst other populations. Similarly, the origins for Himalia and Ananke could be assessed by understanding if Trojans demonstrate surface- variable 3 $\micron$ bands, particularly surrounding impact craters. The Himalia family demonstrates that surfaces with similar colors can nonetheless have different OH-bearing materials, so such searches should not rely only on color parameters.

\begin{acknowledgments}
\section{Acknowledgments}
This work is based on observations made with the NASA/ESA/CSA James Webb Space Telescope. The data were obtained from the Mikulski Archive for Space Telescopes at the Space Telescope Science Institute, which is operated by the Association of Universities for Research in Astronomy, Inc., under NASA contract NAS 5-03127 for JWST. These observations are associated with program \#4028. The specific observations analyzed can be accessed via \dataset[doi: 10.17909/pwcf-c575]{https://doi.org/10.17909/pwcf-c575}. The authors wish to thank James Bauer, Jessica Sunshine, Kathryn Volk, and Shannon MacKenzie for helpful discussions that improved this work. We also thank Driss Takir for providing the original meteorite spectra used for figure comparisons.
\end{acknowledgments}

\bibliography{bibliography}
\bibliographystyle{aasjournal}

\end{document}